\RequirePackage{fix-cm}
\documentclass[smallextended]{svjour3}       
\smartqed  
\usepackage{graphicx}
\usepackage{amsmath}
\usepackage{amsfonts}
\usepackage{graphicx}
\usepackage{color}
\usepackage{enumerate}

\newcommand{\der}[2]{\frac{d#1}{d#2}}

\newcommand{\bc}{\begin{center}}
\newcommand{\ec}{\end{center}}
\newcommand\bea{\begin{eqnarray}}
\newcommand\eea{\end{eqnarray}}
\newcommand\be{\begin{equation}}
\newcommand\ee{\end{equation}}
\newcommand\bi{\begin{itemize}}
\newcommand\ei{\end{itemize}}

\newcommand{\norm}[1]{\vert\vert#1 \vert\vert}

\newcommand\eps{{\varepsilon}}
\newcommand\lam{\lambda}
\newcommand\Lam{\Lambda}

\newcommand\om{\omega}

\newcommand\Ht{{\widetilde H} }

\newcommand\xb{{\overline x} }
\newcommand\yb{{\overline y} }

\newcommand\nb{{\overline n} }

\newcommand\Hb{{\overline H} }

\newcommand\cG{{\cal G} }

\newcommand\gO{{\cal O}}

\newcommand\cD{{\cal D} }

\newcommand\T{\mathbb{T} }

\newcommand\N{\mathbb{N} }

\newcommand\C{\mathbb{C} }

\newcommand\bn{{\bf n} }
\newcommand\bv{{\bf v} }
\newcommand\bvt{\tilde{\bf v} }
\newcommand\bu{{\bf u} }
\newcommand\but{\tilde{\bf u} }

\newcommand{\eexp}{\operatorname{e}}

\newcommand{\dron}[2]{\frac{\partial#1}{\partial#2}}

\newcommand{\qtext}[1]{\quad \text{#1}\quad}
\newcommand\path{{.} }

\begin{document}

\title{Secular Dynamics of Multiplanetary Circumbinary Systems
}
%

\subtitle{Stationary Solutions and Binary-Planet Secular Resonance}

\author{Eduardo Andrade-Ines         \and
				Philippe Robutel
}

\institute{E. Andrade-Ines \at
				IMCCE, Observatoire de Paris - PSL Research University, UPMC Univ. Paris 06, CNRS,
77 Avenue Denfert-Rochereau, 75014 Paris, France \\
Instituto de Astronomia, Geof\'isica e Ci\^encias Atmosf\'ericas (IAG) \\
Rua do Mat\~ao, 1226, Universidade de S\~ao Paulo, S\~ao Paulo, Brazil \\
				\email{eandrade.ines@gmail.com}           			
			 \and
				P. Robutel \at
        IMCCE, Observatoire de Paris - PSL Research University, UPMC Univ. Paris 06, CNRS,
77 Avenue Denfert-Rochereau, 75014 Paris, France \\
				\email{philippe.robutel@obspm.fr}
}

\date{Received: date / Accepted: date}

\maketitle

\begin{abstract}

We present an analytical formalism to study the secular dynamics of a system consisting of $N-2$ planets orbiting a binary star in outer orbits. We introduce a canonical coordinate system and expand the disturbing function in terms of canonical elliptic elements, combining both Legendre polynomials and Laplace coefficients, to obtain a general formalism for the secular description of this type of configuration. With a quadractic approximation of the development, we present a simplified analytical solution for the planetary orbits for both the single planet and the two-planet cases. From the two-planet model, we show that the inner planet accelerates the precession rate of the binary pericenter, which, in turn, may enter in resonance with the secular frequency of the outer planet, characterizing a secular resonance. We calculate an analytical expression for the approximate location of this resonance and apply it to known circumbinary systems, where we show that it can occur at relatively close orbits, for example at $2.4 au$ for the Kepler-38 system. With a more refined model, we analyse the dynamics of this secular resonance and we show that a bifurcation of the corresponding fixed points can affect the long term evolution and stability of planetary systems. By comparing our results with complete integrations of the exact equations of motion, we verified the accuracy of our analytical model.

\end{abstract}

\section{Introduction}
\label{intro}

Binary stars are a frequent phenomenon in the universe, with about 50\% of the known main sequence stars being located at a multiple star system (Abt 1979; Duquennoy \& Mayor 1991; Raghavan et al.  2010, Duch\^ene \& Kraus 2013).  Even though there is still a debate on the planetary formation and protoplanetary evolution in binary star systems, it is expected that many of these systems may host planets (Nelson 2000; Boss 2006; Haghighipour 2006; Th\'ebault et al. 2009; Eggl et al. 2013; Bromley and Kenyon 2015). 

With the Radial Velocity method, the first planets discovered at binary stars from the main sequence were found in the S-type\footnote{Following the classification proposed by Dvorak (1984)} configuration (Eggenberger et al. 2004, 2007; Desidera \& Barbieri 2007; Roell et al. 2012). It was not until recently, with data obtained from the Kepler program, exoplanets in circumbinary orbits\footnote{Also known as P-type, when the planet orbits both stars in an outer orbit} around main-sequence stars have been discovered, being Kepler-16b the first one of them (Doyle et al. 2011). 

The presence of a secondary star induces strong dynamical effects in the planet the smaller is the distance to the binary system. One of the implications of these dynamical effects is that there is a proximity limit in which stable planetary orbits cannot exist (e.g., Dvorak et al. 1989; Holman \& Wiegert 1999; Musielak et al. 2005; Doolin \& Blundell 2011), which implies that the semimajor axis of the stable planetary orbits are much higher than the binary's. Another implication is that the binary is expected to hinder or even halt the planetary formation of the closer orbits, even though they are stable (Moriwaki \& Nakagawa 2004; Meschiari 2012; Paardekooper et al. 2012; Lines et al. 2014). 

For sufficiently distant orbits, planetary formation in circumbinary systems should be similar to the one in single star systems (Bromley and Kenyon 2015), which indicates that multiplanetary circumbinary systems must be very common. However, only one system so far was found in such configuration (the Kepler-47 system, Orosz et al. 2012b, Hinse et al. 2015). This may be due to a selection effect and the presence of additional planets in most of these systems is not discarded (Li et al. 2016).

The fact that most of the currently known exoplanets in circumbinary orbits are close to their binary hosts (e.g., Doyle et al. 2011; Orosz et al. 2012b; Welsh et al. 2012; Schwamb et al. 2013; Kostov et al. 2014; see Table \ref{parameters}), coupled with the fact that outer regions are more friendly for planetary formation (Bromley and Kenyon 2015), suggests that most of these systems must have formed in an outer region and then migrated inwards (e.g., Paardekooper et al. 2012; Marzari et al. 2013; Pierens \& Nelson 2013; Rafikov 2013; Kley \& Haghighipour 2014; Lines et al. 2014; Bromley \& Kenyon 2015; Silsbee \& Rafikov 2015).

Secular dynamics play a major role in the planetary formation and migration processes due to the large time-scales these events usually happen. In the early stages of formation, the secular stationary solutions ({\it i.e.}, the fixed points of the secular problem) are more favourable places in the phase space for the planetary formation due to the smaller collision velocity between the particles that facilitates accretion (Giuppone et al. 2011, Bromley \& Kenyon 2015). Even in the later stages of formation, where smooth migration processes begin to take place, the dynamical evolution is guided for the least energy configurations, {\it i.e.}, the secular stationary solution, due to the dissipative nature of the process (e.g. Michtchenko \& Rodr\'iguez 2011).

The secular dynamics of a planet in a circumbinary orbit have been the subject of study of many authors, more recently by Moriwaki \& Nakagawa (2004), Leung \& Lee (2013), Demidova \& Shevchenko (2015) and  Georgakarakos \& Eggl (2015), among others. Although dealing with the same general problem, these authors adopted different analytical approaches, hypothesis and assumed different approximations in their development, but they all reached consistent and satisfactory results. The only common hypothesis shared between them all was the first-order averaging procedure while constructing the secular part. Differently from the S-type configuration, which requires for many cases the employment of more sophisticated averaging theories for a proper description of the secular dynamics (e.g., Giuppone et al. 2011, Libert \& Sansottera 2013, Andrade-Ines et al. 2016, Andrade-Ines \& Eggl 2017), the first-order averaging theories wield good results to match the integration of the complete equations of motion for the case of planets in circumbinary orbits.

The presence of an additional body, such as a second planet, adds a degree-of-freedom to the system that can significantly change the secular dynamics from the 3-body problem secular solutions. Secular resonances, in particular, are abundant in multi-body systems and have been studied by many authors throughout the years. An example of such effect is the chaotic behaviour of the inner solar system generated by secular resonances (Laskar 1990). 




In this paper we examine some aspects of the secular dynamics of multi-planetary circumbinary systems. One of he goals of this work is to establish a formalism adapted to the construction of the secular Hamiltonian governing the planar motion of planets orbiting a binary star.  We introduce in Section \ref{sec:coordinates} a canonical coordinate system combining elements of both heliocentric and Jacobi coordinates. The Section \ref{sec:expansion} concerns the expansion of the secular Hamiltonian. In order to take into account the different specificities of the problem, we combine expansions in Legendre polynomials and in Laplace coefficients.  Considering the invariance of the total angular momentum, we reduce the problem by one degree of freedom eliminating the coordinates of the binary. 
Once the formalism established, we focus our attention on the existence of relative equilibrium of the system that corresponds fixed points of the reduced secular Hamiltonian.  In Section \ref{sec:quadratic},  we use a simple model where the Hamiltonian is truncated at order 2 in the planetary eccentricities, to give, in the case of one and two planets, an analytic expression of the coordinates of the fixed points as well as their frequencies in the case of stability.  Our analysis highlights the presence of a secular resonance involving the precession frequencies of the binary and the outermost planet, which appears as a singularity in our  previous solutions.   Using higher-order expansions of the secular Hamiltonian, we show in Section \ref{sec:sec_res} that this secular resonance is associated with a bifurcation that gives birth to three families of equilibria, some of which correspond to configurations with high planetary eccentricities.

%
%

\section{The circumbinary coordinates}
\label{sec:coordinates}

\begin{figure}
\begin{center}
\includegraphics[width=0.5\textwidth,angle=0]{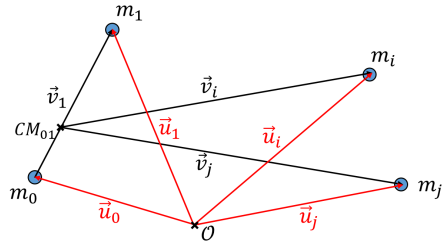}
\caption{Representation of the position vectors $u_i$ with respect to $\cal{O}$ in the inertial reference frame and of the position vectors $v_i$ in the circumbinary canonical coordinate system. $CM_{01}$ stands for the center of mass of the bodies $m_0$ and $m_1$}
\end{center}
\label{def_vec}
\end{figure}

Consider a system composed by two major bodies, of masses $m_0$ and $m_1$, and $n-1$ planets of masses $m_i$, with $2\leq i \leq n$. A natural and symmetric way to build a coordinate system adapted to such planetary systems is to locate the planets with respect to the mass centrer of the binary. As it is represented in Figure \ref{def_vec}, starting from the inertial canonical coordinate system.
$$
(\bu_0, \cdots,\bu_n,\but_0, \cdots , \but_n)  \qtext{with} 
$$
where $\bu_j$ is the position of the $j$-th body with respect to an arbitrary origin {$\cal{O}$ and $\but_j = m_j\dot\bu_j$ } its conjugated momentum. We defined the \textit{circumbinary canonical coordinate system} by the relations
$$
\left\{
\begin{array}{ll}
\bv_0 = & \left(m_0\bu_0 +  m_1\bu_1 + \cdots +m_n\bu_n\right)M^{-1}  \\
\bv_1 = & -\bu_0  + \bu_1 \\ 
\bv_j = & -\eta\bu_0 + (\eta-1) \bu_1 + \bu_j \\
& \\
\bvt_0 = & \but_0 +  \but_1 + \cdots +\but_n  \\
\bvt_1 = &  \but_1 + (1-\eta)(\but_2 + \dots + \but_n) \\
\bvt_j = & \but_j 
\end{array}
\right.
$$

\noindent where  
$$
M = m_0 + \cdots m_n, \quad  m = m_0 + m_1  \qtext{and} \eta = \frac{m_0}{m}.
$$

\noindent In these coordinates, the Hamiltonian function is given by:

\begin{equation}
H = H_0 + H_1,
\label{eq:ham}
\end{equation}
 where 
\be
\begin{split}
&H_0(\bv_j,\bvt_j) = \frac12\frac{\bvt_0^2}{M} +  \sum _{j  = 1}^{n} \left(
    \frac{\bvt_j^2}{2\beta_j} - \frac{\mu_j \beta_j}{\norm{\bv_j}} 
    \right), \\
& \text{with }   
 \beta_1 = \dfrac{m_0m_1}{m},   \mu_1 = \cG m,  \\
& \text{and }  \beta_j = \dfrac{m m_j}{m+m_j},  \mu_j = \cG (m + m_j) \qtext{for} 2\leq j \leq n ,
\end{split}
  \label{eq:ham_H0}
\ee
 is the Keplerian part,  while the perturbation splits in two different terms:

$$
H_1(\bv_j,\bvt_j)      =  H_P(\bv_j,\bvt_j)  + H_B(\bv_j)    
$$
where 
\be
 H_P(\bv_j,\bvt_j) =  \sum_{2\leq j < k\leq n} \frac{\bvt_j\cdot \bvt_k}{m}
-\cG \sum_{2\leq j < k\leq n} \frac{m_j m_k}{\norm{\bv_j - \bv_k}}
\label{eq:ham_HP1}
\ee
and
\be
H_B(\bv_j)  = -\cG \sum_{2\leq j \leq n} m_j   
\left(
\frac{m_0}{\norm{\bv_j - \eta_0\bv_1} }+ \frac{m_1 }{\norm{\bv_j - \eta_1\bv_1} }- \frac{m}{\norm{\bv_j}}
\right), \label{eq:ham_HB1}
\ee
with  $\eta_0 = \eta - 1 = -\dfrac{m_1}{m}$ and  $\eta_1 = \eta = \dfrac{m_0}{m}$, $\cG$ being  the gravitational constant.
These two parts of the Hamiltonian perturbation, namely $H_P$ and $H_B$, describe very distinct subsystems. The \textit{Planetary} part, depending only on the positions and momenta of the planets, describes the motion of the planets perturbed only by each other around the center of mass of the stars 0 and 1. Its expression is the same as  the disturbing Hamiltonian   generated by $n$ planets orbiting a star, expressed in canonical heliocentric coordinates (Laskar \& Robutel, 1995). 
The \textit{Binary} part  $H_B$ describes the circumbinary motion of the $n-1$ planets perturbed only by the binary stars and neglecting the mutual interaction of the planets.  The perturbation (\ref{eq:ham_HB1}) has the same form as a n-body problem expressed in Jacobian coordinates.

%
%

\noindent Henceforth, the first term of the unperturbed Hamiltonian $H_0$, that reflects the fact that the velocity of the center of mass is constant, will be omitted. 

\section{Expansion of the disturbing function}
\label{sec:expansion}

In this section we will present the development of the disturbing function and the domains of its validity.

\subsection{Assumptions and constraints}
\label{sec:constraints}

Let us first list the different assumptions that will be necessary to reduce  the problem and expand the disturbing function.

\begin{enumerate}[(A):]
\item  The masses:  
\be
m_0 \geq m_1     \gg m_2,\cdots ,m_n.
\ee
If we denote by $\eps$ a small parameter reflecting the smallness of the planetary masses with respect to the binary ones, we have $ m_j/m = \gO(\eps)$ for $2\leq j\leq n$.
\label{hypo:H_1}
\item  We consider only the coplanar motion.
\item As we study the secular problem ({\it i.e.}, the problem averaged over the mean longitudes), we have to avoid the major mean-motion resonances.  In order to average the Hamiltonian over the fast angles, we also need to ensure that the largest planetary secular frequency is lower that the mean-motion of the most distant planet. The equality of these two frequencies leads a resonance known as evection resonance (e.g., Touma \& Sridhar 2015). Although this phenomenon is out of the scope of this paper, the location of this resonance is briefly discussed in Section \ref{sec:evec}.
\label{assomp_evec}
\item The planetary eccentricities are sufficiently small to expand the disturbing function in power series in these ones.
\item Stability: the innermost planet has to be far enough of the binary for the system to be stable. We consider stable the orbits that satisfy the criterion derived by  Holman \& Wiegert (1999), that reads
\be
\frac{a_2}{a_1} >  1.60 + 5.10e_1 + 4.12\mu  -2.22e_1^2 - 4.27e_1\mu - 5.09\mu^2 + 4.61e_1^2\mu^2,
\label{eq:crit_bin}
\ee
with $\mu = m_1/m$. For the same stability reason, the distance between two successive planets has to be large enough, and a Hill-like stability criterion (Hill 1878), given by:
\be
\frac{a_{j+1}}{a_j} >  1 + 2.4 \, \left(\frac{\text{max }(m_j,m_{j+1})}{m_0+m_1}\right)^{1/3},
\label{eq:crit_plan}
\ee
is adopted. As the planetary masses are considered small with respect to the total mass of the binary, the ratio $a_j/a_{j+1}$ is not necessarily small and the expansion of the distance between these two planets in terms of Legendre Polynomials would require a large amount of terms. For this reason, we express the planetary part of the disturbing function $H_P$ in terms of Laplace coefficients. On the other hand, we consider the ratios  $a_j/a_1$,for $j\geq 2$, to be sufficiently small such that the binary part $H_B$ is expanded in powers of $a_j/a_1$ by means of Legendre polynomials.
\item 
\label{cond_reduc}
In order to be able to reduce the problem by the elimination of the binary's eccentricity $e_1$ (see the Section \ref{sec:reduction}), we assume that $e_1>0$ and that the binary dominates the largest portion of the total angular momentum of the system. This hypothesis is verified when
\be
 a_n \ll \eps^{-2}a_1.
\ee
This condition is generally satisfied. Indeed, if we have Jupiter-sized planets orbiting Sun-sized stars, the upper bound of the previous inequality is approximately equal $10^6 a_1$, which is too large be of any dynamical interest. 
\label{hypo:reduc}
\item
\label{cond_GR}
We consider only the Newtonian interactions between all the bodies. The influence of post-Newtonian interactions between the two central stars is briefly discussed in Section \ref{sec:PPN}.

\end{enumerate}

\subsection{Keplerian action-angle variables and secular Hamiltonian }

Before going further, we first have to express the Hamiltonian in terms of action-angle variables of the Kepler problem.
Because of the assumptions stated above that regard the eccentricities, Delaunay variables are chosen for the binary while the Poincar\'e variables, which are not singular for zero eccentricity, are more convenient for the planets. 
The Delaunay variables associated with the binary reads
\be
\begin{array}{ll}
 L_1  = \beta_1\sqrt{\mu_1 a_1}, &  M_1, \\[.1cm]
G_1  = L_1\sqrt{1-e_1^2},  & \om_1,
\end{array}
\label{eq:Delaunay}
\ee
while for the $j$th planet, that Poincar\'e variables $(\Lam_j,Q_j,\lam_j,q_j)$ are given by:
\be
\begin{array}{ll}
\Lam_j  = \beta_j\sqrt{\mu_j a_j}, & \lam_j = M_j + \omega_j \, , \\[.1cm]
Q_j  = \Lam_j\left(1-\sqrt{1-e_j^2}\right),  & q_j = -\om_j \, .
\end{array}
\label{eq:Poincare}
\ee 
The following studies being limited to small planetary eccentricities, it is convenient to  introduce the complex canonical Poincar\'e variables $(\Lam_j,\lam_j,x_j,-i\xb_j)$ where: 
$$
x_j =  \sqrt{Q_j}\exp(i\omega_j) = e_j\sqrt{\Lam_j/2}\exp(i\omega_j) (1+ \gO(e_j^2)).
$$
In these variables, the equations of the motion are given by 
$$
\der{x_j}{t} = -i \dron{H}{\xb_j} , \quad \der{\xb_j}{t} =  i \dron{H}{x_j} .
$$
The Keplerian part $H_0$ of the Hamiltonian expressed in terms of action variables reads:
$$
H_0 =  -\frac12\sum_{j=1}^n \dfrac{\mu_j^2\beta_j^3}{\Lam_j^2},
$$
where $\Lambda_1 = L_1$. In the absence of mean motion resonances, the Hamiltonian can be averaged over the fast angles $l_1,\lambda_2, \cdots, \lambda_n$, leading that the secular Hamiltonian $\Hb_1$ is given by the general expression:


\be
\begin{split}
&\Hb(L_1,\Lam_j,G_1,Q_j,g_1,q_j) =  \\
 &\dfrac{1}{ (2\pi)^n}\int_{\T^n } H(L_1,\Lam_j,G_1,Q_j,M_1,\lambda_j,\omega_1,q_j) dM_1d\lambda_2\cdots d\lambda_n
\end{split}
\label{Hsec_int}
\ee
As the secular problem concerns only the long-time variations, the Keplerian Hamiltonian $H_0$, which is an integral of the motion, can be omitted. As a consequence, the secular Hamiltonian reads 
$$
  \Hb_1 = \Hb_P + \Hb_B ,
$$
where the over bar symbol indicates the average over the fast angles.
 \subsection{Expansion of the planetary part}

According to the expression (\ref{eq:ham_HP1}) the planetary part of the disturbing function splits in to two different part. The first, depending only on the planetary  momenta $\bv_j$, reflect the fact that the reference frame associated with the coordinates   $(\bv_2,\cdots,\bv_n,\bvt_2,\cdots,\bvt_n)$ is not Galilean. The $\bvt_j\cdot\bvt_k$ containing only short period terms, they can be discarded. As a consequence, the secular planetary part of the perturbation depends only on the inverse of the mutual distances  $\norm{\bv_j - \bv_k}$ for $2\leq j,k\leq n$.  
In order to get a simple expansion of the secular part of $\norm{\bv_j - \bv_k}^{-1}$, we follow the method developed by LaRo1995. It is based on an development in power series of the inverse of the mutual distance in the Poincar\'e variables $x_j,x_k$ and their conjugated. More precisely, for $2\leq j <k\leq n$, the desired quantity reads:

\be
\Hb_P = -\frac{\cG m_1m_2}{4\pi^2} \int_{\T^2} \frac{a_k\, d\lam_j d\lam_k}{\norm{\bv_j - \bv_k}} = 
\sum_{d>0}\sum_{\bn \in \cD_{2d}} \Gamma_{\bn}(\alpha)  x_j^{n_j}x_k^{n_k} \xb_j^{\nb_j} \xb_k^{\nb_k}  \, ,
\label{eq:devel_plan}
\ee

where $\alpha = a_j/a_k$, $\bn = (n_j,n_k,\nb_j,\nb_k)$  and 
$$
 \cD_{2d} = \left\{   \bn \in \N^4  /  n_j \!+ \!n_k \!+\! \nb_j\!+ \!\nb_k \!= \!2d, n_j \!+ \!n_k = \nb_j\!+\! \nb_k \right\}.
$$
The coefficients $\Gamma_{\bn}(\alpha)$ are explicit functions of $\alpha$ expressed in terms of Laplace coefficients. 
It is shown in LaRo1995 that, if $\bn \in \cD_{2d}$, the simplest expression of $\Gamma_{\bn}$  is of the form 
$$
{\cal P}(\alpha) b_{d+1/2}^{(1)} + {\cal Q}(\alpha) b_{d + 1/2}^{(0)} \, ,
$$
where $b_j^{(k)}$ are Laplace coefficients and  ${\cal P}$ and ${\cal Q}$ are polynomials in $\alpha$ and $\alpha^{-1}$ with rational coefficients. The whole expansion can also be performed using only two Laplace coefficients (see LaRo1995 for more details).
Let us note that, the expression (\ref{eq:devel_plan}) being exact in $\alpha$, it is valid for all $\alpha \in [0,1)$.  
Consequently,  the relative distance between the planets can reach small values.
The sum (\ref{eq:devel_plan}), satisfies specific relations, imposed by the definition of the sets $ \cD_{2d}$, known as D'Alembert rules. The relation linking the different powers of $x_p$ and $\xb_p$ in (\ref{eq:devel_plan}) is equivalent to the fact that the planetary disturbing function depends on the arguments of the perihelia only through their difference, namely $\omega_j - \omega_k$. This is a consequence of the invariant by rotation of the Hamiltonian.
Another direct consequence of this relation is that the total degree of the monomials in $x_p$ and $\xb_p$ is alway even.


\subsection{Expansion of the binary part}

Contrarily to the the mutual distance between the planets, which is not necessary large, we assume that the separation between the stars and the planets (the inner planet) is large enough  so that the quantities $\norm{\bv_j - \eta_k\bv_1} ^{-1}$ appearing in (\ref{eq:ham_HB1}) can be expanded in power series in $\norm{\bv_1}/\norm{\bv_j}$, that is in a series of Legendre polynomials.
An advantage of this development is that is allows us to express the secular part of the problem in terms of the Hansen coefficients (Hansen 1855, Plummer 1918, Kaula 1962, Laskar \& Bou\'e 2010), that are expressions that are exact in the eccentricities for a given degree of the development of the Legendre polynomials. This is a particularly interesting feature of the development as the binary hosting circumbinary planetary systems are known to have arbitrary eccentricities.
The binary part of the disturbing function is rewritten

\begin{equation}
H_B = \displaystyle \sum_{2\leq j \leq n} H_{B_j} \, ,
\label{bin01}
\end{equation}
where
\begin{equation}
H_{B_j}(\bv_j,\bvt_j) =  -\cG m_j   
\left(
\frac{m_0}{\norm{\bv_j - \eta_0\bv_1} }+ \frac{m_1}{\norm{\bv_j - \eta_1\bv_1} }- \frac{m}{\norm{\bv_j}}
\right).
\label{eq:bin02}
\end{equation}
The development of Equation (\ref{eq:bin02}) in Legendre polynomials, after first-order average procedure over the fast angles, can be expressed as (Laskar \& Bou\'e 2010, Andrade-Ines et al. 2016)

\begin{equation}
\Hb_{B_j} = - \frac{\cG m_j}{a_j}\sum_{l=2}^{\infty}\displaystyle\sum_{q=0}^{l} {\cal M}_lf_{l,q} \alpha_j^l X_{0}^{l,2q-l}(e_1) X_{0}^{-l-1,l-2q}(e_j) \eexp^{i(2q-l)\theta_j} \, ,
\label{eq:leg3}
\end{equation}


\noindent where $\alpha_j = a_1/a_j$, $\theta_j = \omega_j - \omega_1$, $X_{a}^{b,c}(e_i)$ are the Hansen coefficients (Hansen 1855) and

\begin{equation}
{\cal M}_l = m_0m_1\frac{m_0^{l-1} - (-m_1)^{l-1}}{(m_0+m_1)^l}, 
\label{massaMl}
\end{equation}

\begin{equation}
f_{l,q} = \frac{(2q)!(2l-2q)!}{2^{2l}((l-q)!)^2(q!)^2}.
\label{fnq}
\end{equation}
According to Laskar \& Bou\'e (2010), the coefficient $X_{0}^{l,2q-l}(e_1)$ is a polynomial of degree $l$ in $e_1$. It is therefore simple to get the exact value of this coefficient whatever the binary eccentricity is. However, it is not the case for the planetary eccentricities. Indeed, one can show that 
 $X_{0}^{-l-1,l-2q}(e_j)$ is a polynomial divided by $\sqrt{1-e_j^2}^{(2l-1)}$. As $e_j$ is assumed to be small for $j\geq 2$, the Hansen coefficients can be expanded in series of $e_j$, leading to the expression 
 \be
X_{0}^{-l-1,l-2q}(e_j) = e_j^{\vert l - 2q\vert} U_{l,q}(e_j^2) \, ,
\label{eq:Hansen}
 \ee
where $U_{l,q}$ is a power series in $e_j^2$. Substituting Equation (\ref{eq:Hansen}) in Equation (\ref{eq:leg3}) we get

\begin{equation}
\Hb_{B_j} = - \frac{\cG m_j}{a_j} \displaystyle\sum_{l=2}^{\infty}\displaystyle\sum_{q=0}^{l} {\cal M}_lf_{l,q} \alpha_j^l X_{0}^{l,2q-l}(e_1)  U_{l,q}(e_j^2) e_j^{\vert l - 2q\vert}\eexp^{i(2q-l)\theta_j}.
\label{eq:leg4}
\end{equation}


\subsection{Reduction of the secular Hamiltonian}
\label{sec:reduction}
As the secular Hamiltonian does not depend on the fast angles, their conjugate actions $L_1, \Lam_2,\cdots ,\Lam_n$ are integrals of motion, which implies that the semimajor axes are constant in the secular dynamics. 
According to (\ref{eq:devel_plan}) and (\ref{eq:leg4}), the disturbing function depends on the angular variables only through $\omega_i - \omega_j$, with $1\leq i,j \leq n$, $i\neq j$. As these quantities can be expressed by the mean of the $n-1$ angular variables $\om_j - \om_1$ for $2\leq j\leq n$, it is natural to introduce the new canonical coordinate system $(\Theta_1,\theta_1,y_2,-i\yb_2,\cdots,y_n,-i\yb_n)$, defined by the relations:
\be
\begin{array}{ll}
\Theta_1 &= G_1 -\left( Q_2 + \cdots + Q_n\right), \\
 \theta_1 &= \omega_1, \\
y_j &= \sqrt{Q_j} \eexp^{-i(\omega_1+q_j)} =  x_j \eexp^{-i\omega_1}\\
&=\sqrt{\beta_j\sqrt{\mu_j a_j}}\sqrt{1-\sqrt{1-e_j^2}} \eexp^{i\theta_j}  \, .
\end{array}
\label{eq:var_reduc}
\ee
As the angle $\theta_1$ does not appear in the disturbing function, its conjugated action $\Theta_1$ is an integral of motion\footnote{Let us notice that  $L_1 - \Theta_1 $  is the angular momentum deficit (Laskar 1997).}.
Through Eqs. (\ref{eq:var_reduc}), we can reduce one of the degrees of freedom of the system by eliminating $G_1$, or the eccentricity $e_1$, substituting this quantity by its expression in terms  of $(\Theta_1, L_1, \vert y_2\vert^2, \cdots, \vert y_n\vert^2)$, that is
\be
e_1 = \left[1-\frac{\Theta_1^2}{L_1^2} - 2\frac{\Theta_1}{L_1} \sum_{j=2}^n \frac{\vert y_j\vert^2}{L_1} - \left(\sum_{j=2}^n \frac{\vert y_j\vert^2}{L_1}\right)^2 \right]^{1/2}.
\label{eq:red_Theta}
\ee 
Considering the hypothesis (\ref{hypo:reduc}), stated in Section \ref{sec:constraints}, we have:
$$
0 < 1-\frac{\Theta_1^2}{L_1^2} <1,
$$
which allows us to introduce the parameter $e_b$ defined by:
\be
e_{b} = \sqrt{1-\frac{\Theta_1^2}{L_1^2}},
\label{eq:e_b}
\ee
and to rewrite the expression (\ref{eq:red_Theta}) in the form:
$$
e_1 = \left[e_b^2 - 2\sqrt{1-e_b^2}\sum_{j=2}^n \frac{\vert y_j\vert^2}{L_1} - \left(\sum_{j=2}^n \frac{\vert y_j\vert^2}{L_1}\right)^2 \right]^{1/2}.
$$
This expression shows that the eccentricity of the binary $e_1$ is $\eps$-close to the constant $e_{b}$ and always smaller.  The equality arises when $y_2 = \cdots = y_n =0$, that is, when the secular planetary orbits are circular or when the planets have negligible masses ($m_j=0$, with $2 \leq j$). In this last case,  the development of $\Hb_{B_j}$ falls to the restricted three-body problem.
The hypothesis (\ref{hypo:reduc}) combined with the expression (\ref{eq:e_b}) leads to the inequality 
$$
e_b^2  >  \left\vert 2\sqrt{1-e_b^2} \sum_{j=2}^n \frac{\vert y_j\vert^2}{L_1} + \left(\sum_{j=2}^n \frac{\vert y_j\vert^2}{L_1}\right)^2 \right\vert ,
$$
which means that the eccentricity $e_1$ can be expanded in a convergent Taylor series in the neighbourhood of $e_b$ as:
\be
e_1 = e_b + \sum_{p=1}^\infty A_p \left(\sum_{j=2}^n \frac{\vert y_j\vert^2}{L_1} \right)^p
= e_b + W\left(e_b,\frac{\vert y_2\vert^2}{L_1},\cdots,\frac{\vert y_n\vert^2}{L_1}\right) \, ,
\label{eq:red_eb}
\ee
where  the coefficients $A_i$ are functions of $e_{b}$. The first three $A_i$ are equal to:
$$
A_1 = -\dfrac{\sqrt{1-e_{b}^2}}{e_{b}} , \quad
A_2 = -\dfrac{1}{2e_b^3}  \qtext{and} 
A_3 = -\dfrac{\sqrt{1-e_{b}^2}}{2e_b^5}.
$$

\subsection{Expression of the Hamiltonian in the new variables}
\label{new_variables}
Let us now focus on the expression of the disturbing function in the variables $y_j$. 
For the planetary part, we just have to replace the variable $x_j$ by $y_j$ in the expression (\ref{eq:devel_plan}). Indeed, as $ n_j \!+ \!n_k = \nb_j\!+\! \nb_k$, we have
 $$
 x_j^{n_j} x_k^{n_k} \xb_j^{\nb_j} \xb_k^{\nb_k}  =  y_j^{n_j} y_k^{n_k} \yb_j^{\nb_j} \yb_k^{\nb_k}.
 $$
As regards the  binary part $\Hb_B$, the calculation is less straightforward. In the expansion (\ref{eq:leg4}) we wish to eliminate the dependence of $e_1$ by performing the reduction of the angular momentum and to transform the variables $(e_j,\theta_j)$ into $(y_j,\yb_j)$. 
The expansion (\ref{eq:leg4}) contains  terms of the form 
\be
 X_{0}^{l,2q-l}(e_1) U_{l,q}(e_j^2) e_j^{\vert l - 2q\vert}\eexp^{i(2q-l)\theta_j}\, ,
 \label{eq:expr_Hansen}
\ee
which can be transformed as follows.
From  (\ref{eq:var_reduc}), we deduce that 
\be
e_j\eexp^{i\theta_j} = y_j \sqrt{\frac{2}{\Lambda_j}\left(1-\frac{\vert y_j \vert^2}{\Lambda_j} \right)}.
\label{eq:transf}
\ee
Therefore, $U_{l,q}(e_j^2)$ can be expressed in function of $\vert y_j \vert^2$ as:  
\be
U_{l,q} (e_j^2) =  U_{l,q} \left(\frac{2\vert y_j \vert^2}{\Lambda_j}\left(1-\frac{\vert y_j \vert^2}{\Lambda_j} \right)\right) = U_{l,q}^\prime(\vert y_j \vert^2), \, 
\label{eq:ulq}
\ee
and the terms $e_j^{\vert 2q-l \vert} \eexp^{i (2q-l)\theta_j}$ give: 
\be
e_j^{\vert 2q-l \vert} \eexp^{i (2q-l)\theta_j} = T_{q,l}(y_j) = 
\left\{
\begin{array}{rcl}
\yb^{(l-2q)} \left( \displaystyle\frac{2}{\Lambda_j} - \displaystyle\frac{\vert y_j \vert^2}{ \Lambda_j^2}\right)^{(l-2q)/2} & \textrm{if} & q<\frac{l}{2}\, ,\\
y^{(2q-l)} \left( \displaystyle\frac{2}{\Lambda_j} - \displaystyle\frac{\vert y_j \vert^2}{ \Lambda_j^2}\right)^{(2q-l)/2}  & \textrm{if} & q\geq \frac{l}{2}\, .\\
\end{array}
\right.
\label{eq:hanplan1}
\ee
Using the relation (\ref{eq:red_eb}), we eliminate the dependence of $e_1$ of (\ref{eq:expr_Hansen}), and remembering that the function $W$ is of order $\gO (\eps)$, we get

\be
X_{0}^{l,2q-l}(e_b+W(\vert y_j \vert^2)) = X_{0}^{l,2q-l}(e_b) + F_{l,q}(\vert y_j \vert^2),
\label{eq:exp_han_e1}
\ee
where $F_{l,q}$ is a power series in $\vert y_j \vert^2$ and is of order $\gO (\eps)$. This last remark will lead to a natural splitting of the Binary Part of the disturbing function (\ref{eq:leg4}). Introducing (\ref{eq:ulq}), (\ref{eq:hanplan1}) and (\ref{eq:exp_han_e1}) into (\ref{eq:leg4}) we get

\be
{\widetilde H}_{B_j} = \Ht_{B_{j}}^{(b)} + \Ht_{B_{j}}^{(r)},
\label{red-eq7}
\ee
where the tilde denotes that the Hamiltonian is expressed in the reduced variables (\ref{eq:var_reduc}), and

\begin{equation}
\Ht_{B_j}^{(b)} = - \frac{\cG m_j}{a_j} \sum_{l=2}^{\infty}\sum_{q=0}^{l} {\cal M}_lf_{l,q} \alpha_j^l X_{0}^{l,2q-l}(e_b)  U_{l,q}^\prime(\vert y_j \vert^2) T_{l,q}(y_j),
\label{eq:restrict}
\end{equation}
is the \textit{main} part, that is of order $\gO(\eps)$, and

\begin{equation}
\Ht_{B_j}^{(r)} = - \frac{\cG m_j}{a_j} \sum_{l=2}^{\infty}\sum_{q=0}^{l} {\cal M}_lf_{l,q} \alpha_j^l F_{l,q}(\vert y_j \vert^2)  U_{l,q}^\prime(\vert y_j \vert^2) T_{l,q}(y_j),
\label{eq:reduced}
\end{equation}
is the \textit{reduced} part, that is of order $\gO(\eps^2)$.

 At this point, we note that the \textit{main} part acquires a similar form to the secular Hamiltonian of the restricted three-body problem if we take $e_b = e_1$, while the \textit{reduced} part contains the terms that describe the motion induced by the planets on the central binary. Indeed, developing the Hamiltonian (\ref{eq:leg4}) in a Taylor series around $e_1 = e_b$ and using (\ref{eq:red_eb}) we have 
\be
\begin{array}{rl}\vspace{0.4cm}
\Hb_1(e_1,y_j,\yb_j) = &\Hb_1\left(e_b+W(|y_j|^2),y_j,\yb_j\right)\\ \vspace{0.4cm}
= &\Hb_1(e_b)(e_b,y_j,\yb_j) + W(|y_j|^2)\displaystyle\frac{\partial \Hb_1}{\partial e_1}(e_b,y_j,\yb_j) + \gO(W^2).
\end{array}
\label{eq:ex_red1}
\ee
By the substitution of the expression  (\ref{eq:red_eb}) in (\ref{eq:ex_red1}) and using the relation 
$$
\frac{\partial \Hb_1}{\partial e_1} = {\frac{\partial \Hb_1}{\partial \Theta_1}}\frac{\partial \Theta_1}{\partial e_1} = -\dot{\omega}_1 \frac{L_1 e_1}{\sqrt{1-e_1^2}},
$$
we get

%
$$
\Hb_1(e_1,y_j,\yb_j) = \Hb_1(e_b,y_j,\yb_j) + \dot{\omega}_1 \sum_{i=2}^n |y_i|^2 + \gO(|y_j|^4).
$$
In this last equation we identify the first term $\Hb_1(e_b,y_j,\yb_j)$ to be the main part of the Hamiltonian (\ref{red-eq7}), while the rest, that is order $\gO(\eps)$, is the reduced part of (\ref{red-eq7}). This last term, or at least its main part (term in $|y_j|^2$), can be interpreted as an inertial term which comes from the fact that the reduced coordinate system rotates with the pericenter of the binary, whose precession rate is equal to $\dot\omega_1$. 
This is the reason why the precession rate of the binary can be deduced from the reduced part of the secular Hamiltonian (see sections \ref{sec:np_1} and \ref{sec:np_2}).

Finally, the secular Hamiltonian of the coplanar problem of $n-2$ planets in circumbinary orbits is written as
\be
\Ht_1 = \Ht_P(y,\yb) + \sum_{j=1}^n \left(\Ht_{B_j}^{(b)}(y,\yb) + \Ht_{B_j}^{(r)}(y,\yb)\right),
\label{eq:sechamiltotal}
\ee
where $\Hb_P$, $\Hb_{B_j}^{(b)}$ and $\Hb_{B_j}^{(r)}$ are given by (\ref{eq:devel_plan}), (\ref{eq:restrict}) and (\ref{eq:reduced}), respectively.
The binary parts are of order one in the planetary masses and their expressions in power series of  $(y,\yb)$ contain all possible degrees greater or equal to one, for the main part $\Hb_{B_j}^{(b)}$, and to two, for the reduced expression $\Hb_{B_j}^{(r)}$.  As in the classical planetary problem, the planetary part  $\Hb_P$ is of order two in the planetary masses and contains only monomials of even total degrees in  $(y,\yb)$.

\section{Quadratic approximation}
\label{sec:quadratic}

The secular Hamiltonian (\ref{eq:sechamiltotal}) is expanded in a power series of  $(y,\yb)$, and as a consequence, its main dynamical features (at least for small eccentricities) can be deduced from the study of its parts of degree one and two, that is, the linear approximation of its associated differential system.  As mentioned in the previous section, the Hamiltonian contains linear terms and, therefore, circular planetary orbits are not solution of the secular problem, opposed to the case of the usual secular planetary problem. As we will see later, the solutions of the system oscillate around fixed points which correspond to ellipses precessing at the same rate as the binary's apsidal line. As a degeneracy associated with a secular resonance arises when at least  two planets orbit the binary (Sections \ref{sec:np_2} and \ref{sec:sec_res}), the case of a single planet will be studied separately (Section \ref{sec:np_1}).
 
From this point forward, it is convenient to replace the planetary masses $m_j$ by $\eps m_j$ ($2\leq j$), to identify the most important terms in the complete Hamiltonian. This leads to replace the variables $y_j$  by $\sqrt{\eps} y_j$  and to divide the Hamiltonian (\ref{eq:h_single1}) by $\eps$ in order to keep the canonical form of the equations of the motion. Note that we can always find the original equations simply by taking $\eps = 1$. 


\subsection{A single planet ($n=2$)}
\label{sec:np_1}
The configuration with a single planet has been extensively studied (e.g., Moriwaki \& Nakagawa, 2004) and it will serve as a benchmark to our model and see if we can reproduce the classical results. Naturally, in this case the planetary part vanishes and the expression of the secular Hamiltonian (\ref{eq:sechamiltotal}), limited to its terms of degree lower or equal to $2$, reads

\be
\Ht_1 = \Ht_{B_2}^{(b)} + \eps \Ht_{B_2}^{(r)},
\label{eq:h_single1}
\ee
with 
$$
\Ht_{B_2}^{(b)} = B_2^{(1)} (y_2 + \yb_2) + B_2^{(2)}y_2 \yb_2,
$$
and
$$
\Ht_{B_2}^{(r)} = R^{(2)} y_2 \yb_2,
$$
where the coefficients $B_2^{(j)} $  and $R^{(2)} $ depend on the masses, the semimajor axes and the eccentricity parameter $e_b$.
For  conciseness, theirs expressions are truncated at degree $3$ in  $\alpha_2 = a_1/a_2$, which gives:  
$$
B_2^{(1)} = -\frac{3}{16} n_1 \frac{m_0 m_1 (m_0 - m_1)}{m^3} \sqrt{2\Lambda_2}\alpha_2^{9/2}\left[-\frac{5}{2}e_b - \frac{15}{8}e_b^3 \right],
$$
$$
B_2^{(2)} = -\frac{3}{4}n_1\frac{m_0 m_1}{m^2} \alpha_2^{7/2} \left[1+ \frac{3}{2}e_b^2 \right],
$$
$$
R^{(2)} = \frac{3}{4}n_1 \frac{m_2}{m}\alpha_2^3\sqrt{1-e_b^2},
$$
where $n_1 = \sqrt{\cG m/a_1^3}$ is the mean motion of the central binary. 
As the secular frequency
\be
g_2 = - (B_2^{(2)} + \eps R^{(2)})
\label{eq:freq_single}
\ee
is strictly positive in the domain where the angular momentum reduction is valid\footnote{The equation $-(B_2^{(2)} + \eps R^{(2)}) =0$ has an unique solution when $\alpha = \gO(\eps^2)$, which does not satisfy the Hypothesis \ref{hypo:reduc}. } the solution of the canonical equation associated with the Hamiltonian (\ref{eq:h_single1}) reads:
\be
y_2(t) = \eexp^{ig_2 t} \beta_{2}  + \,  y_{2}^{(f)}, 
\label{eq:sol_single}
\ee
where $\beta_2 \in \mathbb{C}$ is an integration constant depending on the planetary secular initial conditions and
$$
y_{2}^{(f)} = \frac{B_2^{(1)}}{g_2}.
$$
It turns out that the solution (\ref{eq:sol_single}) rotates with the frequency $g_2$ around a stable fixed point of coordinates $y_{2}^{(f)}$.  This  equilibrium corresponds to an elliptic planetary orbit whose eccentricity is equal to:  
$$
e_{2}^{(f)} = \sqrt{\frac{2}{\Lambda_2}} \vert y_{2}^{(f)}\vert,
$$
and whose pericenter is aligned  with the binary's one.
As mentioned in the Section \ref{sec:reduction}, the precession rate of the binary can be deduced from the reduced part. Thus, the secular frequency associated with the binary is approximated by:
$$
g_1 = \eps R^{(2)}.
$$
By neglecting the terms of order $\gO(\eps)$ in (\ref{eq:h_single1}), we find the expression given in  Moriwaki \& Nakagawa (2004) for the forced eccentricity and the secular frequency in to the restrict problem's approximation.

\subsection{Two planets ($n=3$)}
\label{sec:np_2}

Let us now consider the general case, which will be illustrated  by a system including two planets orbiting a binary star. As in the previous section, the binary part will be truncated to order $3$ in $\alpha_j = a_1/a_j$, while the expression in terms of Laplace coefficients will allow us to keep exact expressions in  the semimajor axis ratio of the planets $a_2/a_3$.
According to Section \ref{sec:reduction}, the expression of the secular Hamiltonian (\ref{eq:sechamiltotal}), limited to its terms of degree lower or equal to $2$ in $(y,\yb)$, reads
\be
\Ht_1 = \Ht_{B}^{(b)} + \eps \Ht_{B}^{(r)} + \eps \Ht_P,
\label{cp-eq0}
\ee
where 
\be
\Ht_{B}^{(b)} = B_2^{(1)}(y_2 + \yb_2) +  B_3^{(1)}( y_3 + \yb_3) +  B_2^{(2)} y_2 \yb_2 + B_3^{(2)} y_3 \yb_3,
\label{HBb}
\ee
\be
\Ht_{B}^{(r)} =  R^{(2)}( y_2 \yb_2 + y_3 \yb_3),
\label{HBr}
\ee
\be
H_P = P^{(1)}( y_2 \yb_3 + \yb_2 y_3 ) + P^{(2)}_2 y_2 \yb_2 + P^{(2)}_3 y_3 \yb_3,
\label{HPp}
\ee
with the coefficients approximated by
$$
B_j^{(1)} = -\frac{3}{16} n_1 \frac{m_0 m_1 (m_0 - m_1)}{m^3} \sqrt{2\Lambda_j}\alpha_j^{9/2}\left[-\frac{5}{2}e_b - \frac{15}{8}e_b^3 \right],
$$
\be
B_j^{(2)} = -\frac{3}{4}n_1\frac{m_0 m_1}{m^2} \alpha_j^{7/2} \left[1+ \frac{3}{2}e_b^2 \right],
\label{bj2}
\ee
\be
R^{(2)} = \frac{3}{4}n_1 \sqrt{1-e_b^2}\left(\frac{m_2}{m} \alpha_2^3+ \frac{m_3}{m}\alpha_3^3 \right),
\label{eq:r2}
\ee
$$
P^{(1)} = -2n_1\frac{\sqrt{m_2m_3}}{m} \alpha_2^{1/4} \alpha_3^{5/4} C_2\left(\frac{\alpha_3}{\alpha_2}\right),
$$
$$
P^{(2)}_2 = -n_1\frac{m_3}{m} \alpha_2^{1/2} \alpha_3 C_3\left(\frac{\alpha_3}{\alpha_2}\right),
$$
$$
P^{(2)}_3 = -n_1\frac{m_2}{m} \alpha_3^{3/2} C_3\left(\frac{\alpha_3}{\alpha_2}\right),
$$
 where $n_1 = \sqrt{\cG m/a_1^3}$ is the mean motion of the central binary and
$$
C_2(\alpha) = \frac{3}{8}\alpha b_{3/2}^{(0)}(\alpha) - \frac{1}{4}(1+\alpha^2)b_{3/2}^{(1)}(\alpha)  
 \quad = -\frac{\alpha}{8}b_{3/2}^{(2)}(\alpha),
$$
$$
C_3(\alpha) = \frac{1}{4}\alpha b_{3/2}^{(1)}(\alpha) ,
$$
where $b_{s}^{(k)}(\alpha)$ are the Laplace coefficients (Laskar \& Robutel, 1995). The coefficients $C_2$ and $C_3$ are exact with respect to the semimajor axis ratio $\alpha$ while expressed in terms of the Laplace coefficients, but if $\alpha$ is sufficiently small, they acquire the approximate expressions
$$
C_2(\alpha) = -\frac{15}{16}\alpha^3 - \frac{105}{64}\alpha^5 - \frac{4725}{2048}\alpha^7 + \gO(\alpha^9)
$$
$$
C_3(\alpha) = \frac{3}{4}\alpha^2 + \frac{45}{32} \alpha^4 + \frac{525}{256}\alpha^6 + \gO(\alpha^8).
$$

As in the Poincar\'e cartesian variables the equations of motion of the Hamiltonian (\ref{cp-eq0}) are given by

\be
\frac{d y_j}{dt} = -i \frac{\partial \Ht_1}{\partial \yb_j} {, \ \rm} \frac{d \yb_j}{dt} = i \frac{\partial \Ht_1}{\partial y_j} \qtext{with} j\in \{2,3 \}
\label{eq-motion}
\ee
and because the Hamiltonian $ \Ht_1$ is a real function, the canonical system (\ref{eq-motion})  can be reduced to a differential system in $\C^2$ where the unknown vectorial function $Y$ satisfies:
\be
\frac{dY}{dt}  =    \frac{d}{dt}  \left(\begin{array}{c}
y_2 \\
y_3 \\
\end{array}\right) =
-i\left( {\cal A} Y +  {\cal B} \right)
\label{eqdif}
\ee
with
$$
{\cal A} =
\left(\begin{array}{cc}
 B_2^{(2)} + \eps P^{(2)}_2 + \eps R^{(2)}& \eps P^{(1)} \\
\eps P^{(1)} & B_3^{(2)}  +\eps P^{(2)}_3 +\eps R^{(2)} \\
\end{array}\right)
\text{and} \,\,
{\cal B} = 
\left(\begin{array}{c}
B_2^{(1)} \\
B_3^{(1)} \\
\end{array}\right).
\label{matA}
$$
If the eigenvalues of $-{\cal A}$, given by
\be
g_j =  -\left( B_j^{(2)} + \eps P^{(2)}_j + \eps R^{(2)} \right)+ \gO (\eps^2){\rm ,\ } j \in \{2,3\}
\label{eq:eigenvalues}
\ee
are different from 0, the solution of Equation (\ref{eqdif}) reads
\be
Y = \beta_2 \eexp^{ig_2 t} {\cal V}_2 + \beta_3 \eexp^{ig_3 t} {\cal V}_3 +  Y_F,
\label{eq:sol2plan}
\ee
where $\beta_i \in {\mathbb C}$ are constants of integration depending on the planetary initial conditions, ${\cal V}_j$ are the eigenvectors of the matrix ${\cal A}$ and $Y_F$ is the stationary solution. Denoting $y_{j}^{(f)}$ the two  coordinates of this particular solution, one has:
\be
y_{2}^{(f)} = \frac{B_2^{(1)}}{g_2} - \frac{\eps P^{(1)} B_3^{(1)}}{g_2 g_3} + \gO(\eps^2),
\quad  y_{3}^{(f)} = \frac{B_3^{(1)}}{g_3} - \frac{\eps P^{(1)} B_2^{(1)}}{g_2 g_3} + \gO(\eps^2),
\label{eq:y_j}
\ee
while the eigenvectors read:
$$
{\cal V}_2 = \zeta \left( 
\begin{array}{c}
g_2 - g_3 \\
\eps P^{(1)}
\end{array} \right) {\rm , \ } {\cal V}_3 = \zeta \left( 
\begin{array}{c}
\eps P^{(1)}\\
g_3 - g_2
\end{array} \right)
$$
with
$$
\zeta = \left((g_2 - g_3)^2 + (\eps P^{(1)})^2 \right)^{-1/2}.
$$
Similarly to the case of a single planet, the solutions (\ref{eq:sol2plan}) oscillate around the stable fixed points $Y_F$ following a quasiperiodic motion with frequencies $g_2$ and $g_3$.  In terms of orbital elements, these stationary solutions correspond to two ellipses of eccentricities\be
e_{j}^{(f)} = \sqrt{\frac{2}{\Lambda_j}} | y_{j}^{(f)} |,
\label{ejf}
\ee
whose pericenter is aligned or anti-aligned with the binary one, depending whether the sign of $y_{j}^{(f)}$ is positive or negative, respectively.
The perturbations generated by the planets on the the binary make its pericenter precess with a secular frequency approximated by:
$$
g_1 = \eps R^{(2)} = \frac{3}{4}n_1 \eps \sqrt{1-e_b^2}\left(\frac{m_2}{m} \alpha_2^3+ \frac{m_3}{m}\alpha_3^3 \right).
$$

\subsubsection{ Location of the evection resonance}
\label{sec:evec}
According to the assumption (\ref{assomp_evec}), the highest secular frequency must be smaller than the smaller mean-motion, which corresponds to: 
\be
n_3 \gg g_2.
\label{eq:evec1}
\ee
As a consequence, the equality of these two frequencies, which coincides with an evection resonance, provides a bound of validity of our secular model.
Neglecting the terms of higher order in $\eps$, the evection resonance will happen when the semimajor axis ratio $\alpha_3$ reaches a critical value $\alpha_E$ such that
\be
\alpha_E^{3/2} = \frac{3}{4} \frac{m_0 m_1}{m^2} \alpha_2^{2} \left[1 + \frac{3}{2}e_b^2 \right].
\label{eq:evec2}
\ee
Therefore, the secular model is no longer valid if $\alpha_3 < \alpha_E$.

\section{Secular resonance with the binary}
\label{sec:sec_res}

The stationary solutions presented in the previous section are valid if $g_j\neq 0$,  what has been assumed thus far. In this section, we will consider the dynamical implications of such degeneracy, associated with a secular resonance.


\subsection{Location of the secular resonance}
Let us fix the values of the semimajor axes $a_1$ and $a_2$, letting $a_3$ vary. If the outer planet is close enough to the inner one, the quantity $B_j^{(2)}$ dominates the expression (\ref{eq:eigenvalues}) of the secular frequencies, which are consequently positive. By increasing $a_3$, the ratio $\alpha_3=a_1/a_3$ begins to decrease, and we have 
\be
\begin{array}{rl}
\vspace{0.4cm}
g_3 =&  -(B_3^{(2)} + \eps R^{(2)} + \eps P_3^{(2)})  +\gO(\eps^2) \\
\vspace{0.4cm}
= &  {\displaystyle\frac{3}{4}}n_1\left(\displaystyle\frac{m_0 m_1}{m^2} \alpha_3^{7/2} \left[1+ \displaystyle\frac{3}{2}e_b^2 \right] - \eps \sqrt{1-e_b^2}\displaystyle\frac{m_2}{m} \alpha_2^3
\right)  \\
 +& \, \gO(\alpha_3^{9/2},\eps\alpha_3^3,\eps^2)\, ,
\end{array}
\label{eq:g3-resso}
\ee
where the term of planetary interactions $ \eps P_3^{(2)} $, which is of order $\eps\alpha_3^3$, is rejected in the remainder\footnote{Of course, this approximation in not valid if the planets are close one to another (see Figures \ref{gs_kep38-fo} and \ref{ef_kep38-fo})}. Following this approximation, the frequency $g_3$ vanishes when $\alpha_3$ reaches a critical value $\alpha_R$ given by:
\be
\alpha_R^{7/2} =  \eps\frac{mm_2}{m_0 m_1} \alpha_2^3 \sqrt{1-e_b^2} \left[1+ \frac{3}{2}e_b^2 \right]^{-1} \, .
\label{eq:alp-SR}
\ee
As a consequence, when $\alpha_3\gg \alpha_R$, the secular frequency $g_3$ is dominated by the term $-B_3^{(2)}$, that is strictly positive. The frequency $g_3$ becomes negative only when  $\alpha_3\ll \alpha_R$ where $-\eps R^{(2)}$ dominates.

\subsection{Orbital interpretation}

At this point, it is interesting compare the influence, on the secular frequencies and forced eccentricities, of the different parts of the Hamiltonian (\ref{cp-eq0}), namely the Binary Main part (\ref{HBb}), the Binary Reduced part (\ref{HBr}) and the Planetary part (\ref{HPp}). The frequencies and the location of the fixed point are obtained by applying the same method used to construct the general solution for each part of the Hamiltonian. This leads to 
\be
g_{j}^{(b)} = - B_j^{(2)}, {\rm \ } e_{j}^{(b)} = \sqrt{\frac{2}{\Lambda_j}} \frac{B_j^{(1)}}{g_{j}^{(b)}},
\label{g_b}
\ee
for the Binary Main part, where $g_{j}^{(b)}$ corresponds to the precession frequency of each planet induced by the binary, measured in a fixed reference frame.

The Binary Reduced part, whose contribution is the same for both planets, gives
\be
g^{(r)} = - \eps R^{(2)}, {\rm \ } e^{(r)} = 0,
\label{g_r}
\ee
where $-g^{(r)}$ is the precession rate of the binary.
Finally, the planetary interactions generate the terms 
\be
g_{j}^{(p)} =  \displaystyle\frac{\eps}{2} \left[ P_2^{(2)} + P_3^{(2)} + (-1)^j \sqrt{(P_2^{(2)} - P_3^{(2)})^2 + 4(P^{(1)})^2}\right], \quad e_{j}^{(p)} =  0,
\label{g_p}
\ee
where the $g_{j}^{(p)} $ are the secular frequencies of the two planets orbiting a central body of mass $m_0+m_1$.

\begin{figure}
\begin{center}
\includegraphics[width=0.8\textwidth,angle=0]{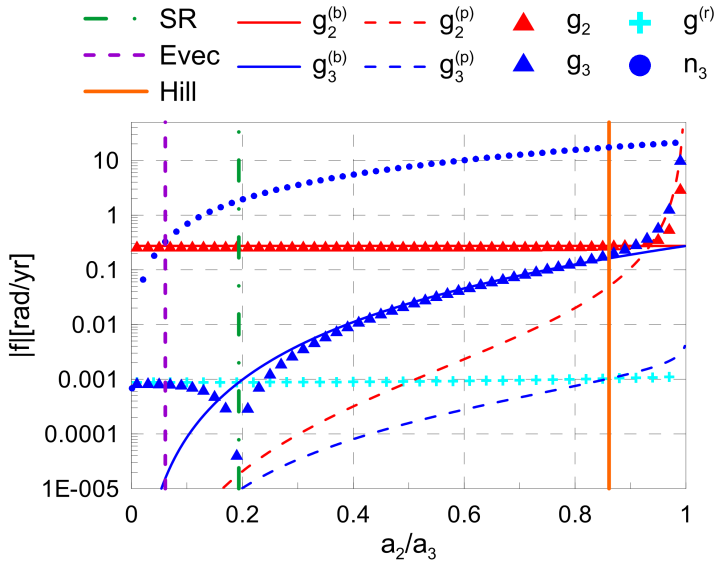}
\caption{Secular frequency in function of $a_2/a_3$ calculated with the different models for the Kepler-38 system. The curves in blue stand for the inner planet 2, and the curves in dark blue stand for the outer planet 3. The continuous lines represent the Binary Main part ($g_j^{(b)}$), the dashed lines represent the Planetary part ($g_j^{(p)}$) and the triangles represent the complete secular model ($g_j$). The light blue crosses represent the Binary Reduced part ($g^{(r)}$) for both planets and the dark blue circles represent the mean motion for the second planet ($n_3$). The vertical lines in green and purple show the position for the secular resonance (SR, Eq. \ref{eq:alp-SR}) and for the evection resonance (Evec, Eq. \ref{eq:evec2}), respectively. The vertical line in orange shows the Hill stability limit, given by Equation (\ref{eq:crit_plan}). } 
\label{gs_kep38-fo}
\end{center}
\end{figure}

We present at Figure \ref{gs_kep38-fo} the secular frequencies calculated with the complete secular model (Equation \ref{eq:eigenvalues}), as well as the partial secular frequencies (Eqs. \ref{g_b} to \ref{g_p}) for the system Kepler-38 (Table \ref{parameters}) composed by an additional fictitious planet of mass $m_2 = 10^{-4} M_\odot$ and with semimajor axis varying in the range $a_3 \in [0.487,46.4]$ (or simply $a_2/a_3 \in [0.01,0.99]$). Figure \ref{gs_kep38-fo} shows as well the approximate position of the secular resonance (vertical green line), given by Equation (\ref{eq:alp-SR}), the position for an evection resonance (vertical purple line), given by Equation (\ref{eq:evec2}), and the planetary stability limit (vertical orange line), given by Equation (\ref{eq:crit_plan}).

Figure \ref{gs_kep38-fo} portraits a complex behavior of the the secular frequency for the complete secular model (red and blue triangles) that can be separated into 5 intervals regarding the semimajor axis ratio $a_2/a_3$. 
We note that at $a_2/a_3 \approx 0.05$ the condition for the evection resonance is satisfied. As it was presented in Section \ref{sec:evec}, the model is valid only if $a_2/a_3 > \alpha_E a_2/a_1 \approx 0.05$. Therefore, the dynamical behaviour for $a_2/a_3 < 0.05$ may not be well described by our model for this particular case.

For $0.05<a_2/a_3 <0.1$, the complete secular Hamiltonian is dominated by the Binary Reduced parts: the main part for $g_2$ ($g_2 \approx g_2^{(b)}$) and the reduced part for $g_3$ ($g_3 \approx g^{(r)}$). As mentioned in the Section \ref{new_variables}, the Binary Reduced term $g_j^{(r)}$is the result of a transformation of coordinates from the complete secular problem and is equivalent to the precession on the central binary induced by the planets, or in this case, the first planet. Due to the large distance of the secondary planet, it does not have any dynamical influence in the system, which means that the pericenter of the binary evolves in a constant rate due only to the first planet, that causes the precession of the reduced angle $\theta_3 = \omega_3 - \omega_1$ in the opposite sense.

The region of the space of parameters in the interval $0.1 < a_2/a_3 < 0.3$ contains the secular resonance, located at $a_2/a_3 \approx 0.2$ for this particular example. We notice that the resonance is located exactly at the intersection of the $g_j^{(r)}$ (magenta) and the $g_3^{(b)}$ (blue full) curves, and that is the same value of $a_2/a_3$ that the complete secular frequency $g_3$ (blue triangles) reaches the value 0. 

For $0.3 < a_2/a_3 < 0.85$ we note that the complete secular model is in good accordance with the results obtained only by the Binary Main part: the triangles curve match very well the full curves. As the Binary Main part has the same form as the restricted 3-body problem (considering $e_1 = e_b$), another  possible interpretation is that the restricted approximation properly describes the secular behavior of the system for this region of the space of parameters.

Finally, for $ a_2/a_3 > 0.85 $ we note that the planetary frequencies $g_j^{(p)}$ increase drastically, which also increases the values of the frequencies obtained by the complete secular model. However, we note that this happens in a unstable region of the space of parameters, for semimajor axis ratio higher than the Hill stability limit adopted (Equation \ref{eq:crit_plan}), represented as the vertical orange lines in Figure \ref{ef_kep38-fo}, at $a_2/a_3 \approx 0.86$. It is worth noting that the location of the stability limit is a function of the individual masses of the planets: smaller planetary masses could shift the instability limit to larger values of $a_2/a_3$, where the Planetary part could be a good approximation to the problem. We emphasize, however, that as $a_2/a_3$ increases, mean motion resonances between the planets become more and more important and the averaging procedure performed in the development of this model loses its validity.

\subsection{Location of the fixed points}
\label{sec:location}

\begin{figure}
\begin{center}
\includegraphics[width=0.8\textwidth,angle=0]{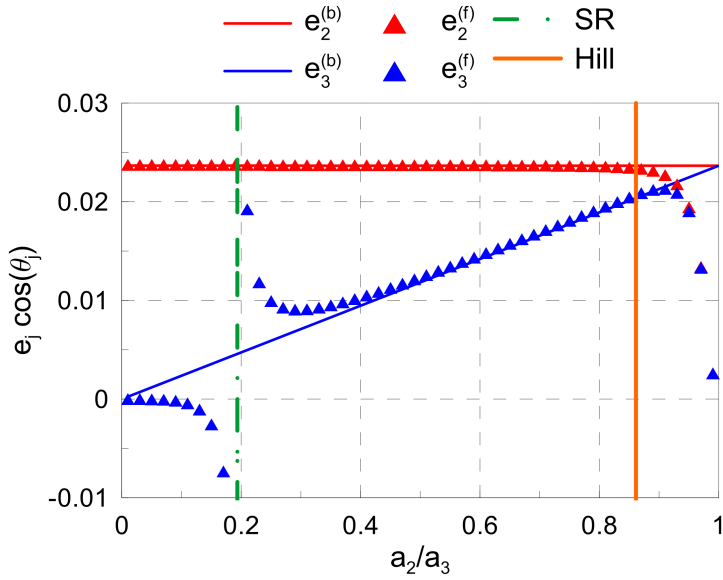}
\caption{Same as Figure \ref{gs_kep38-fo}, but for the forced eccentricity. Note that for this case we omitted the forced eccentricities for the Planetary and Binary Restricted parts of the Hamiltonian, as $e_j^{(p)} = e_j^{(r)} = 0$.} 
\label{ef_kep38-fo}
\end{center}
\end{figure}

The position of the fixed points are also greatly affected by the secular resonance. We present at Figure (\ref{ef_kep38-fo}) the value of the forced eccentricity of the fixed point in function of the semimajor axis ratio $a_2/a_3$ for the same system portrayed in Figure \ref{gs_kep38-fo}.

From Equations (\ref{eq:y_j}) and (\ref{ejf}), straightforward calculations show that when $g_3$ tends to zero (at the resonance), the two forced eccentricities $e_{j}^{(f)}$ behave as:

\be
e_2^{(f)} = \gO\left( g_3^{-1} \eps^{24/7}\right) \qtext{and}
e_3^{(f)} = \gO\left( g_3^{-1} \eps^{9/7}\right)
\label{width-res}
\ee

The singularity is visible, at least for $e_3^{(f)}$, in Fig. \ref{ef_kep38-fo} where the curve gets closer to its vertical asymptote. As  $e_2^{(f)}/e_3^{(f)} = \gO\left(\eps^{15/7}\right)$, this singularity is difficult to perceive for $e_2^{(f)}$ due to a poor resolution on the  x-axis.We note also that, as we cross the secular resonance, the fixed point shifts its plane at the figure: for $\alpha>0.2$ it is located at $\theta_3 = 0$ and for $\alpha<0.2$ it is located at $\theta_3 = \pi$. 

We note again a good agreement between the results obtained by the complete model and the Binary Main part for $0.3 < a_2/a_3 < 0.85$. For greater values of the semimajor axis ratio, at $ a_2/a_3 > 0.85 $ , we note that both the forced eccentricities drop to 0, which agrees with the result obtained from Figure \ref{gs_kep38-fo}, that the model approaches the Planetary solution. This would mean as well that the problem, in this region, could be approximated as the planetary problem, revolving around a star with mass $m=m_0+m_1$. However, this effect happens in unstable region in the space of parameters.

As $\alpha_3$ approaches 0, from Equations (\ref{ejf}) and (\ref{g_b}) one can show that
\be
\frac{e_3^{(f)}}{e_3^{(b)}}\sim 2\eps^{-1} \frac{m_0m_1}{mm_2} \frac{1+\frac32e_b^2}{\sqrt{1-e_b^2}} \alpha_2^{-3} \alpha_3^{7/2},
\label{e3f-lim}
\ee
where we see that $e_3^{(f)}$ decreases faster than $e_b^{(b)}$ as $\alpha_3$ decreases. This result agrees with the behavior observed for $a_2/a_3 < 0.1$ on Figure \ref{ef_kep38-fo}. Another conclusion drawn from Equation (\ref{e3f-lim}) is that the Binary Main part\footnote{Recall that the Binary Main part can be interpreted as the restricted 3-body problem approximation, with $e_b = e_1$.} is not a good approximation for the system in this range of parameters. This also means that even far from the secular resonance, there can still be significant effects in the planetary orbits induced by the indirect effect of a inner planet present in the system.

\subsection{Application to known systems}

We present at Table \ref{parameters} the physical and orbital parameters of known circumbinary star systems, as well as the semimajor axi given by Equation (\ref{eq:alp-SR}), of the secular resonance and by Equation (\ref{sec:sec_res}) of the evection resonance, considering a hypothetical second planet of mass $m_3 = 10^{-4} M_\odot$ in a outer orbit.

\begin{table}
\caption{Parameters of circumbinary systems and the position of the secular resonance and of the evection resonance. We assumed for our simulations all orbits to be coplanar and the planetary masses to be minimal within the limits given by the authors.}
\label{parameters}
\begin{tabular}{llllllllll}
\hline
System & $m_0$ & $m_1$ & $m_2$ & $a_1$ & $e_1$ & $a_2$ & $e_2$ & $a_{3R}$ & $a_{3E}$ \\ 
 & $[M_\odot]$ & $[M_\odot]$ & $[10^{-4} M_\odot]$ & $[au]$ & - & $[au]$ & - &  $[au]$ & $[au]$ \\ \hline
Kep-16b$^a$ & 0.687 & 0.202 & 3.15 & 0.224 & 0.16 & 0.72 & 0.024 & 3.64 & 12.9 \\
Kep-34b$^b$ & 1.049 & 1.022 & 2.10 & 0.228 & 0.521 & 1.086 & 0.209 & 9.34 & 21.2 \\
Kep-35b$^b$ & 0.885 & 0.808 & 1.24 & 0.176 & 0.142 & 0.605 & 0.048 & 5.24 & 9.41 \\
Kep-38b$^c$ & 0.949 & 0.249 & 3.63 & 0.147 & 0.103 & 0.464 & 0.032 & 2.40 & 8.57 \\
Kep-47b$^c$ & 1.043 & 0.362 & 0.296 & 0.084 & 0.023 & 0.296 & 0.035 & 3.34 & 5.79 \\
Kep-47c$^c$  & 1.043 & 0.362 & 0.687 & 0.084 & 0.023 & 0.99 & 0.41 & 7.40 & 96.8 \\
Kep-64b$^d$ & 1.528 & 0.378 & 5.07 & 0.174 & 0.212 & 0.634 & 0.054 & 3.36 & 14.0 \\
Kep-413b$^e$ & 0.82 & 0.542 & 2.00 & 0.099 & 0.037 & 0.355 & 0.118 & 2.45 & 6.1 \\
Kep-1647b$^f$ & 1.22 & 0.97 & 14.51 & 0.13 & 0.16 & 2.72 & 0.058 & 9.70 & 471 \\
\hline
\end{tabular}
{\scriptsize  Notes: References of the systems data: $^a$Doyle et al. (2011); $^b$Welsh et al. (2012);  $^c$Orosz et al. (2012a); $^d$Schwamb et al. (2013), Kostov et al. (2013); $^e$Kostov et al. (2014); $^f$Kostov et al. (2016).}
\end{table}

Table \ref{parameters} shows that, the semimajor axis of the evection resonance is larger than the one of the secular resonance. This implies the dynamics of the secular resonance is within the bounds of applicability of our secular model for all these systems. Additionally, we note that the secular resonance can be found in relatively close orbits for the circumbinary multiplanetary systems. For those systems, the secular resonance can be found as close as $a_3 = 2.4 au$ for the Kepler-38 system and even in the farthest case, for the Kepler-1647 system, the resonance happens at $a_3 = 9.7au$, which is of the order of Saturn’s semimajor axis in the Solar System.
This illustrates the relevance of the secular resonance, that may play an important part in the history of formation and evolution of circumbinary multiplanetary systems. 

\subsection{Bifurcation generated by the  secular resonance}
\label{higher_order}



As shown in the previous section, the dynamics near the Secular Resonance may drastically increase the values of the eccentricities of the planets, in particular of the second planet in the outer orbit. However, the quadratic approximation has a limited range of validity in terms of the eccentricities, compromising the results for this region of the space of parameters. Thus, to properly understand the influence of the Secular Resonance to larger values of eccentricities, a higher-order model is required.

Generally speaking, resonances are associated with changes in the topology of the phase space, which are identified as bifurcations. In Section \ref{rough} we present a first extension of the quadratic model to a higher-order theory in which we can identify the bifurcation associated with the Secular Resonance. A more sophisticated model capable of quantitatively reproducing the numerical integrations of the exact equations of motion is presented in Section \ref{general}.

\subsubsection{A rough model}
\label{rough}

Let us first consider a very simple approximation of the secular Hamiltonian 
(\ref{eq:sechamiltotal}) which allows us to get rid of the singularity arising from the 
quadratic approximation studied in Section \ref{sec:quadratic}.  In order to build this 
approximation, let us assume that the eccentricity of the first planet $e_2$ is small 
enough to be neglected. Then we set $y_2=0$. As a consequence, we are left with a one 
degree of freedom Hamiltonian depending on the variable $y_3$ and of its conjugated 
variable. The two last approximations consist in truncating the resulting Hamiltonian at 
degree $3$ in $\alpha_j$ and $4$ in $y_3,\yb_3$.
Denoting by $F_4$ this function, a stationary $(y_3^{(0)}, \yb_3^{(0)})$ satisfies the 
polynomial equation system:
$$
\dron{F_4}{y_3}(y_3^{(0)}, \yb_3^{(0)}) = \dron{F_4}{\yb_3}(y_3^{(0)}, \yb_3^{(0)}) = 0 .
$$
As the stationary solutions that we have found in Section (\ref{sec:np_2}) are symmetric, in the sense that their apsidal lines are collinear to the one of the binary, we assume that the solutions that we seek in this section have the same symmetry. Imposing  that $y_3^{(0)} = \yb_3^{(0)}$, we only have to 
find the real roots of a third degree polynomial whose coefficients are real numbers. More 
precisely, this polynomial equation reads:
\be
Q(X) = \sum_{j=0}^3 c_jX^j = 0,
\label{eq:poly_Q}
\ee
where the coefficients $c_j$ are given by
\be
\begin{split}
128 m c_0  &= 15{m_0m_1(m_0-m_1)}(3e_b^3 +4e_b) \alpha_3^{9/2} \\
16c_1   &= -3m_0m_1(3e_b^2 +2) \alpha_3^{7/2}  + 6\eps m_2m\sqrt{1-e_b^2}\alpha_2^3\\
1024 m c_2 &= 855 m_0m_1(m_0-m_1)(3e_b^3 + 4e_b)\alpha_3^{9/2} \\
8 c_3   &= - 3m_0m_1(3e_b^2+2)\alpha_3^{7/2}.
\end{split}
\ee
For a given root of this equation, the corresponding eccentricity is deduced from the relation:
$e_3^{(f)} = X\sqrt{1-X^2/4}$.
The quadratic approximation of the secular Hamiltonian used in (\ref{eq:sechamiltotal}) 
leads to singular solutions in the neighbourhood of the resonance, that is, for $\alpha_3$ close to $\alpha_R$. This higher-order approximation allows us to get rid of this problem. 
Indeed, as $\alpha_R = \gO(\eps^{2/7})$, in this neighbourhood the size of the coefficients $c_j$ verifies the relations:
$$
c_0 = \gO(\eps^{9/7}),\, c_1 = \gO(\eps),\, c_2 = \gO(\eps^{9/7}),\,  c_3 = \gO(\eps).
$$
As a result, in the neighbourhood of the resonance, the discriminant of  the polynomial $Q$, given by (\ref{eq:poly_Q}), satisfies:
$$
\Delta = -4c_1^3c_3 +  \gO(\eps^{32/7}) = \gO(\eps^4).
$$
The coefficient $c_3$ being negative, the sign of $\Delta$ is the same as the one of $c_1$. Therefore, $Q$ possesses one  real root when $\alpha_3> \alpha_R$  and three real  roots for $\alpha_3 <\alpha_R$. Therefore, under the hypothesis stated in this section, 
for a given $\alpha_3$ greater than $\alpha_R$, the Hamiltonian system possesses only one (symmetric, when the planets and binary have aligned or anti-aligned apsidal lines) equilibrium. When  $\alpha_3=\alpha_R$, a new real double root emerges from 
the complex plane and bifurcates to give birth to two additional equilibria when $\alpha$ is smaller that $\alpha_R$.

\subsubsection{General case}
\label{general}

The simple model provided in the previous section gives a qualitative description of what we expect that will happen in the secular resonance for a simplified example.
Let us now consider the general Hamiltonian (\ref{eq:sechamiltotal}). Once the parameters 
$m_j, e_b, a_j$ of the studied system are given, the Hamiltonian that drives the secular motion of the planets does not depend any longer on anything but the four variables $(y_2,\yb_2,y_3,\yb_3)$.  Denoting by $F$ this Hamiltonian, the equilibria of its associated canonical equations are solution of the polynomial system:
\be
\begin{split}
\dron{F}{y_2}(y_2, \yb_2,y_3, \yb_3) = \dron{F}{\yb_2}(y_2, 
\yb_2,y_3, \yb_3)  =  0, \\
\dron{F}{y_3}(y_2, \yb_2,y_3, \yb_3)  = \dron{F}{\yb_3}(y_2, 
\yb_2,y_3, \yb_3) = 0,
\end{split}\ee
which can be solved numerically. For each solution, the dynamical nature of the corresponding fixed point is deduced from the eigenvalue of the Hessian matrix of the Hamiltonian $F$ evaluated at this point. If the real part of the four eigenvalues are equal to zero, the fixed point is  elliptic and therefore, the equilibrium is (at least linearly) stable. If, on the contrary, one of the real parts is different from zero, the point is hyperbolic and consequently, unstable. 

\begin{figure}
\begin{center}
\includegraphics[width=0.8\textwidth,angle=0]{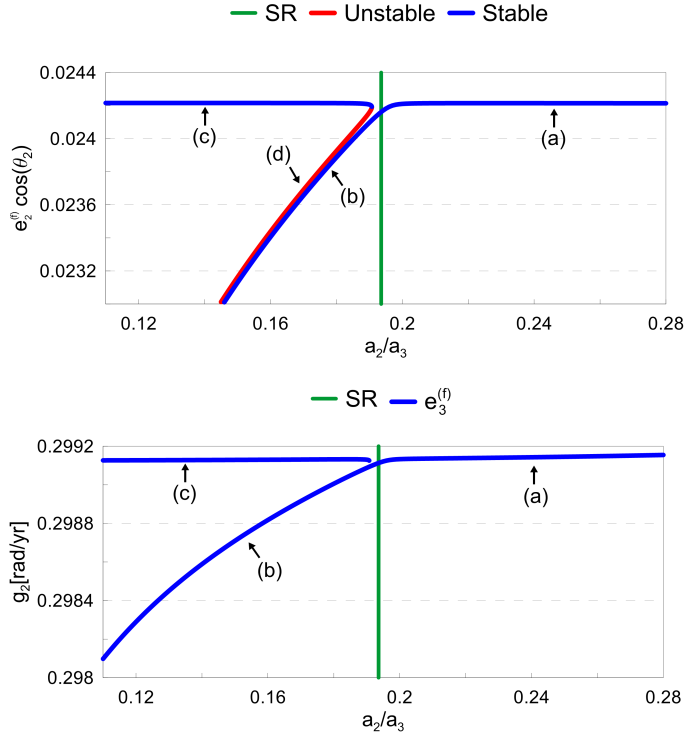}
\caption{Eccentricity of the fixed-point (top) and its frequency (bottom) of the first planet ($j=2$) in function of the semimajor axis ratio $a_2/a_3$ calculated with the high-order model (10 in $e$ and 6 in $\alpha_j$) for the Kepler-38 system. The curve in blue display the position of the elliptic (stable) fixed points, while the curve in red displays the position of the hyperbolic (unstable) ones. The vertical dashed green curve represents the position of the Secular Resonance (SR) for the system. Note that the frequencies for the hyperbolic fixed points (branch (d) in red in the top panel) are not presented in the bottom panel.} 
\label{fixed_point-gs-HO-p2}
\end{center}
\end{figure}

\begin{figure}
\begin{center}
\includegraphics[width=0.8\textwidth,angle=0]{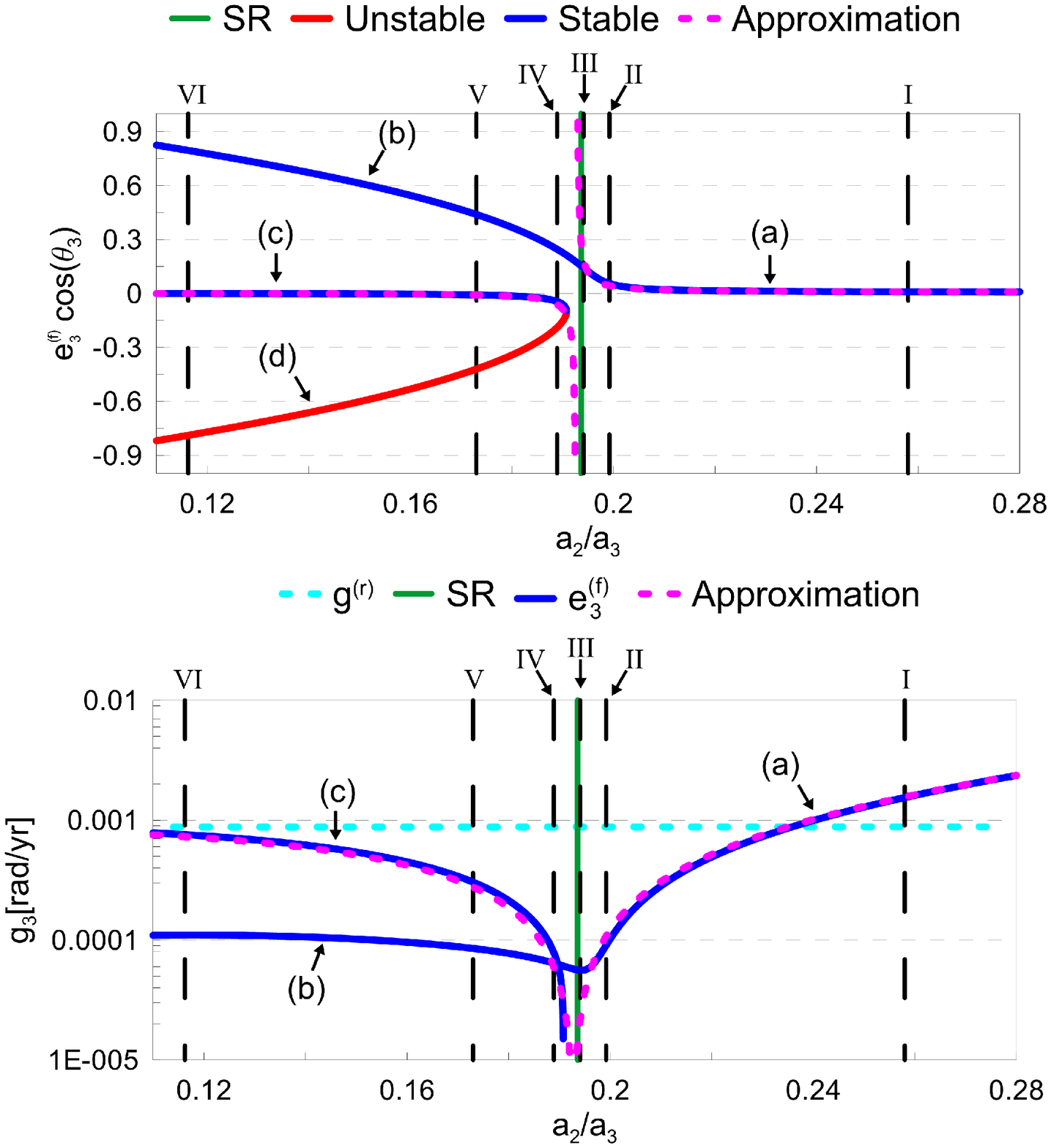}
\caption{Eccentricity of the fixed-point (top) and its frequency (bottom) of the second planet ($j=3$) in function of the semimajor axis ratio ($a_2/a_3$) calculated with the high-order model for the Kepler-38 system. The curve in blue display the position of the elliptic (stable) fixed points, while the curve in red displays the position of the hyperbolic (unstable) ones. The vertical dashed green curve represents the position of the Secular Resonance (SR) for the system, while the vertical dashed lines in black are associated with a Roman numeral (I to VI), each corresponding to a series of integrated orbits for different initial values of $e_3 \cos \theta_3$, that are presented at Figure \ref{levels}. As in Figure \ref{fixed_point-gs-HO-p2}, the eigenvalues of the unstable family (d) are not presented in the bottom panel.}
\label{fixed_point-gs-HO-p3}
\end{center}
\end{figure}

We applied this method to the case of the system Kepler-38 whose parameters $m_0, m_1, m_2$, $a_1,a_2$ and $e_1$ are gathered in Table \ref{parameters}. The mass of the outer planet is again set to $m_2 = 10^{-4} M_\odot$, while its semimajor axis $a_3$ varies such that $a_2/a_3 \in [0.11:0.28]$. For this system, we found that all fixed points correspond to symmetric equilibria. We present the results obtained, with the Hamiltonian (\ref{eq:sechamiltotal}) truncated at degree $10$ in the planetary eccentricities and $6$ in the $\alpha_j$, for the forced eccentricity and the secular frequencies of these fixed points for the inner planet at Figure \ref{fixed_point-gs-HO-p2} and for the outer planet at Figure \ref{fixed_point-gs-HO-p3}. To better analyse the curves, we also denote each branch of the curves for the fixed points by the letters (a) to (d). It is also worth emphasizing that the results depicted in both Figures \ref{fixed_point-gs-HO-p2} and \ref{fixed_point-gs-HO-p3} correspond to two different projections of the equilibrium on the planes $(e_j\cos\theta_j, e_j\sin\theta_j)$, for $j=2$ in Figure \ref{fixed_point-gs-HO-p2} and $j=3$ in Figure \ref{fixed_point-gs-HO-p3}. 

The top panels of both Figures \ref{fixed_point-gs-HO-p2} and \ref{fixed_point-gs-HO-p3} show the bifurcation phenomenon that was predicted in Section \ref{rough}.  As it was suggested by our simplified model, the system possesses only one fixed point for $\alpha_3 > \alpha_R$, that is when $a_2/a_3 > \alpha_R a_2/a_1 \approx 0.193$,  corresponding to the branch denoted by the label (a),  while for $\alpha_3 < \alpha_R$, or $a_2/a_3 < \alpha_R a_2/a_1 \approx 0.193$, there exist three fixed points (branches (b), (c) and (d)). The branch (a)-(b) corresponds to the same continue family of stable fixed points, while the families (c), which contains elliptic fixed points, and (d), composed of unstable equilibria, arise from the bifurcation at $\alpha_3 = \alpha_R$.

Even though the bifurcation is visible in the variables of both planets, the influence of the secular resonance is much stronger in the second planet (Figure \ref{fixed_point-gs-HO-p3}) than it is in the first one (Figure \ref{fixed_point-gs-HO-p2}). Note that the scale on the $y-$axis on both figures is not the same, so the difference between the forced eccentricity (and therefore for the secular frequency as well) of the different branches of the first planet (Figure \ref{fixed_point-gs-HO-p2}) is much smaller than the one of the second planet (Figure \ref{fixed_point-gs-HO-p3}), which agrees with the result presented at Section \ref{sec:location} (Equation \ref{width-res}). Indeed, for the first planet, the forced eccentricities along the families (b) and (d) differs in value by a few percent from branches (a) and (b), and even less for the secular frequency $g_2$, which is not significant.

Differently of the first planet, we note from Figure \ref{fixed_point-gs-HO-p3} that the forced eccentricity of all three fixed points are very distinct for the second planet, reaching values as high as $\approx 0.9$. Additionally, the fixed points of the branches (a) and (b) have all the apsidal lines aligned to binary's ($\theta_3 = 0$), while the branches (c) and (d) have their fixed points with apsidal lines anti-aligned to the binary ($\theta_3 = \pi$).
It also worth noticing that outside of a given neighbourhood of the resonance, represented by the green vertical line in  Figure \ref{fixed_point-gs-HO-p3}, the quadratic approximation (presented in Section \ref{sec:np_2}), is able to accurately describe the branches (a) and (c) of the stable equilibria.  
Even though the expressions of $e^{(f)}$ and $g$ derived from the quadratic approximation of the secular Hamiltonian are not valid around the bifurcation, the location of this latter is correctly predicted by this approximation (Equation \ref{eq:alp-SR}). 

\begin{figure}
\begin{center}
\includegraphics[width=0.32\textwidth,angle=270]{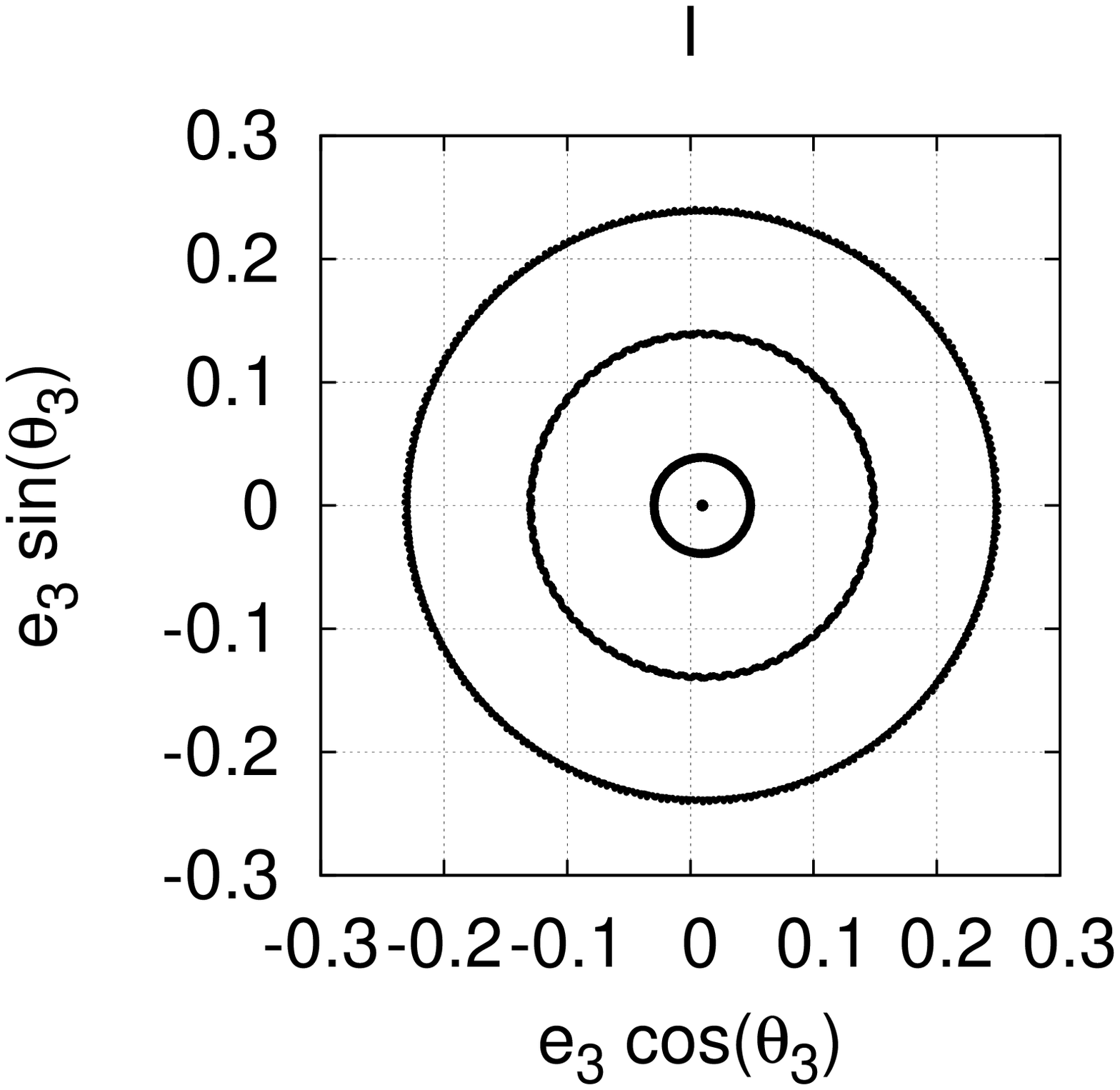}
\includegraphics[width=0.32\textwidth,angle=270]{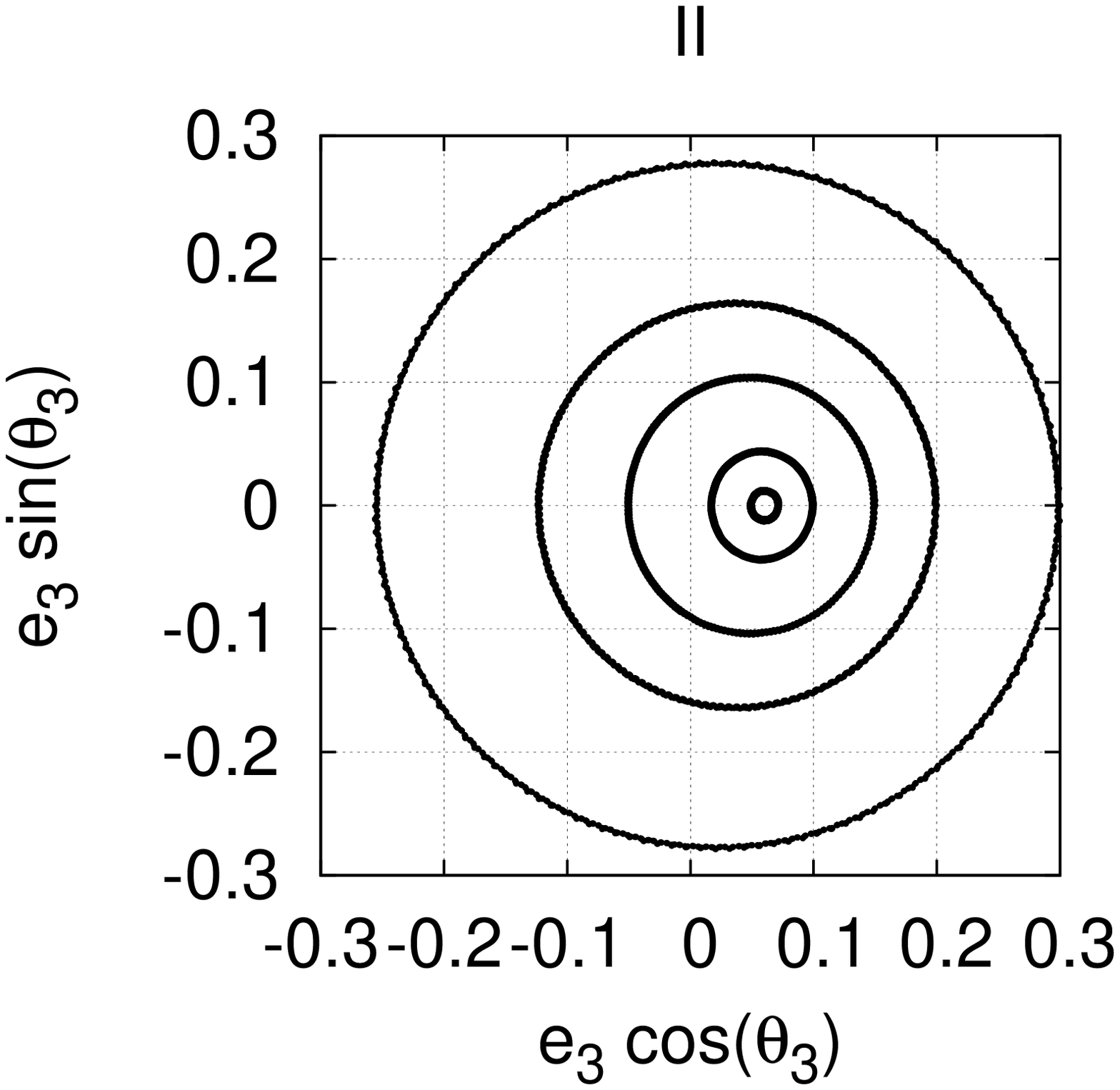}
\includegraphics[width=0.32\textwidth,angle=270]{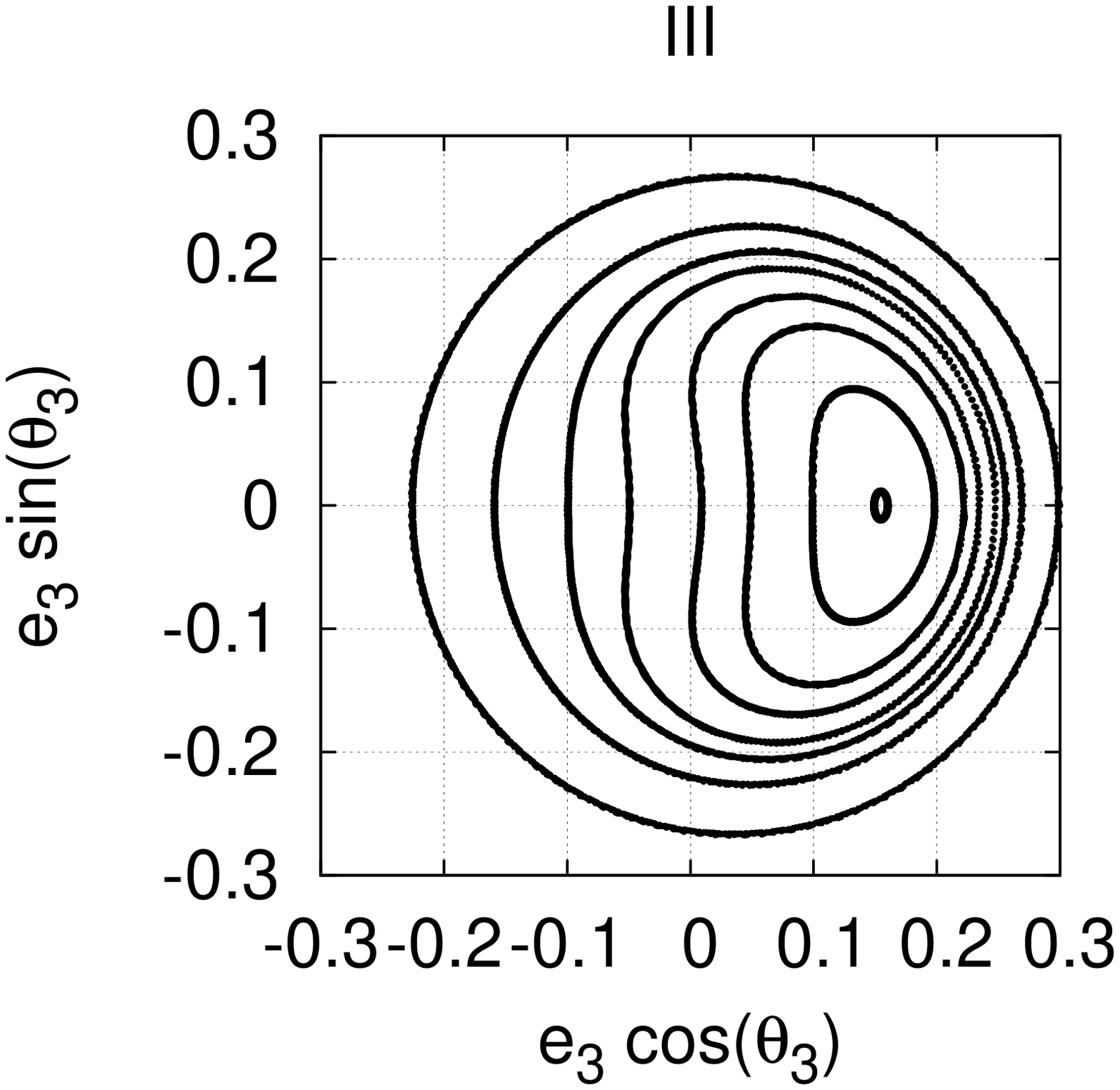}
\includegraphics[width=0.32\textwidth,angle=270]{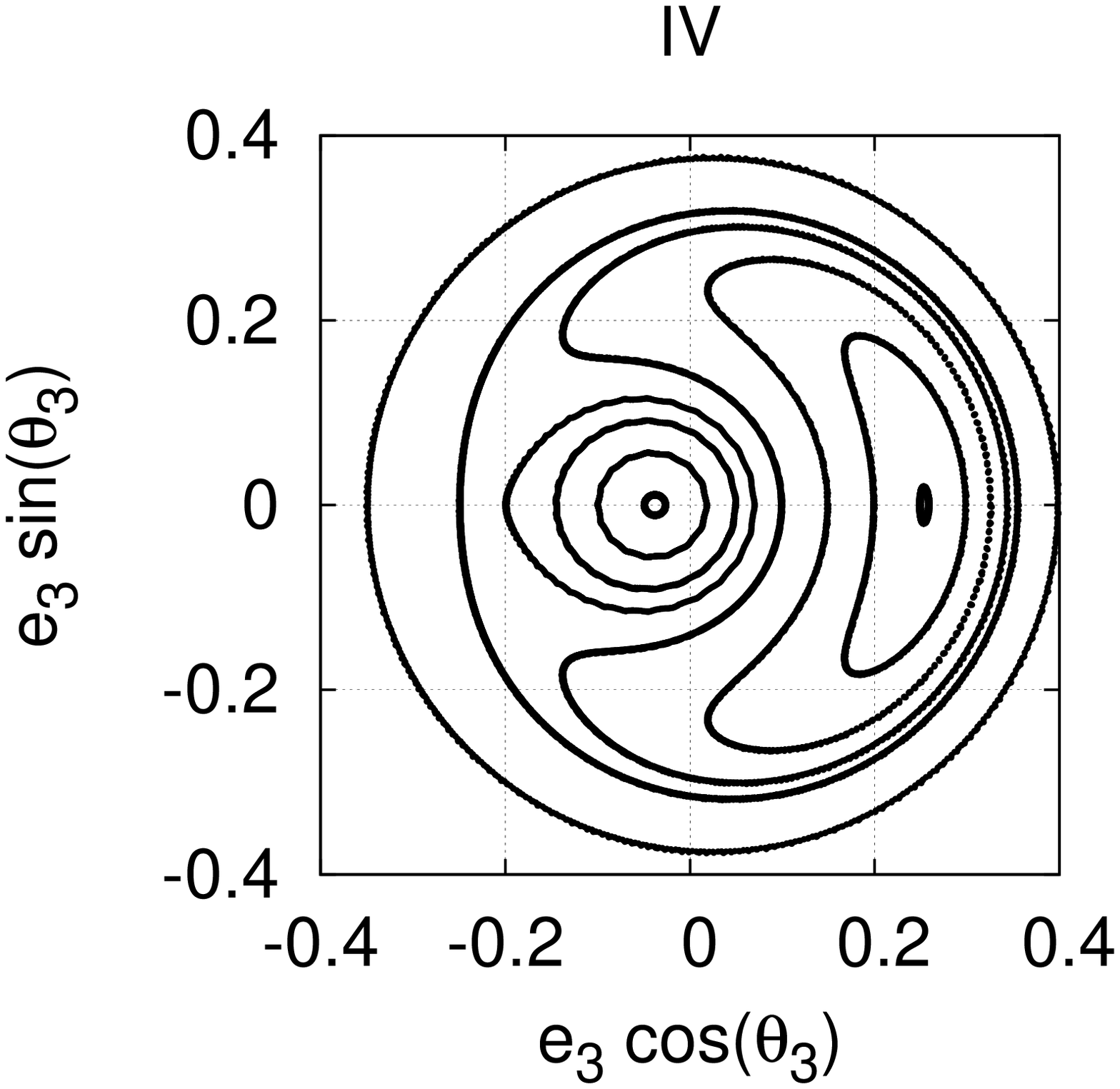}
\includegraphics[width=0.32\textwidth,angle=270]{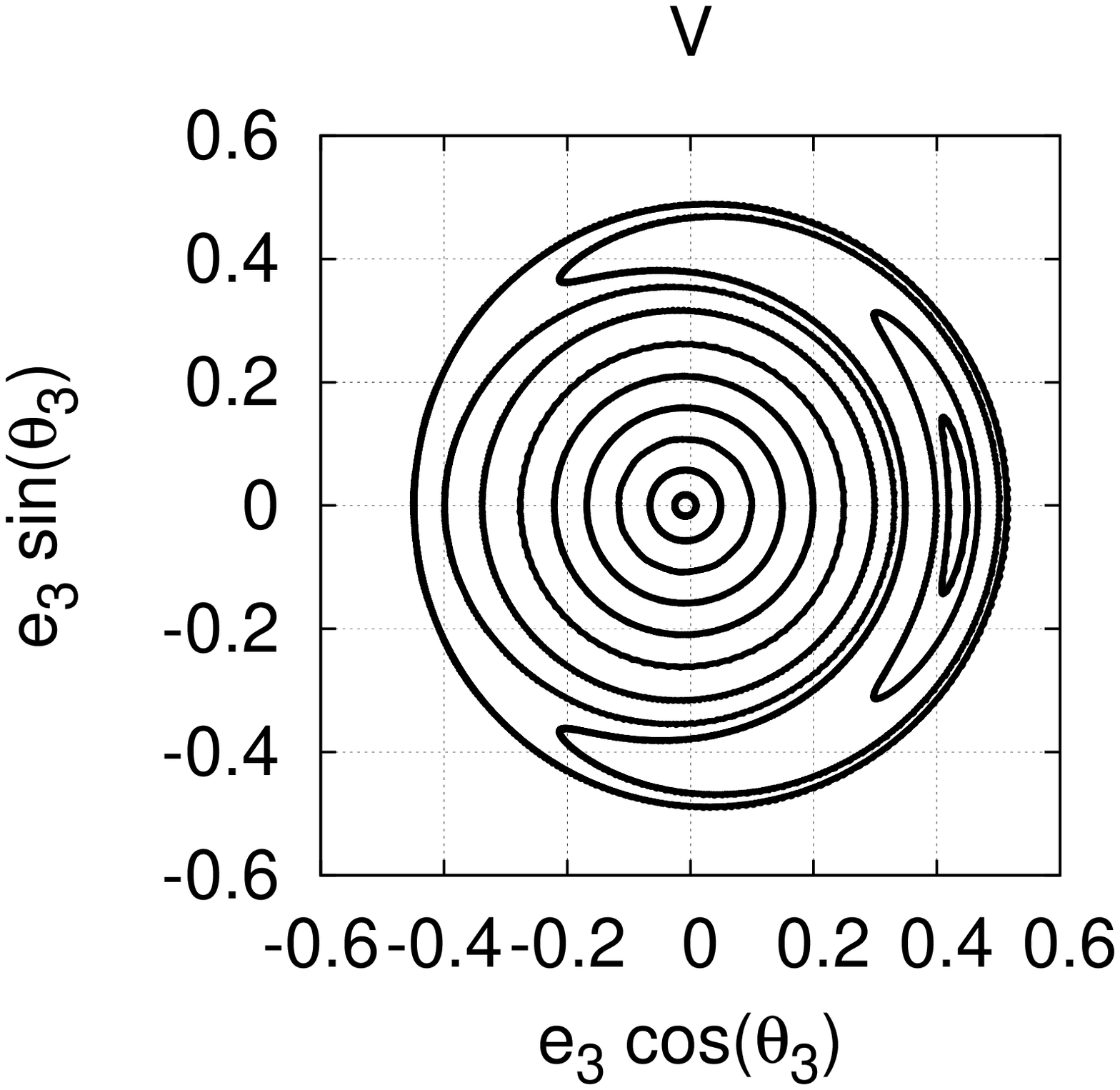}
d\includegraphics[width=0.32\textwidth,angle=270]{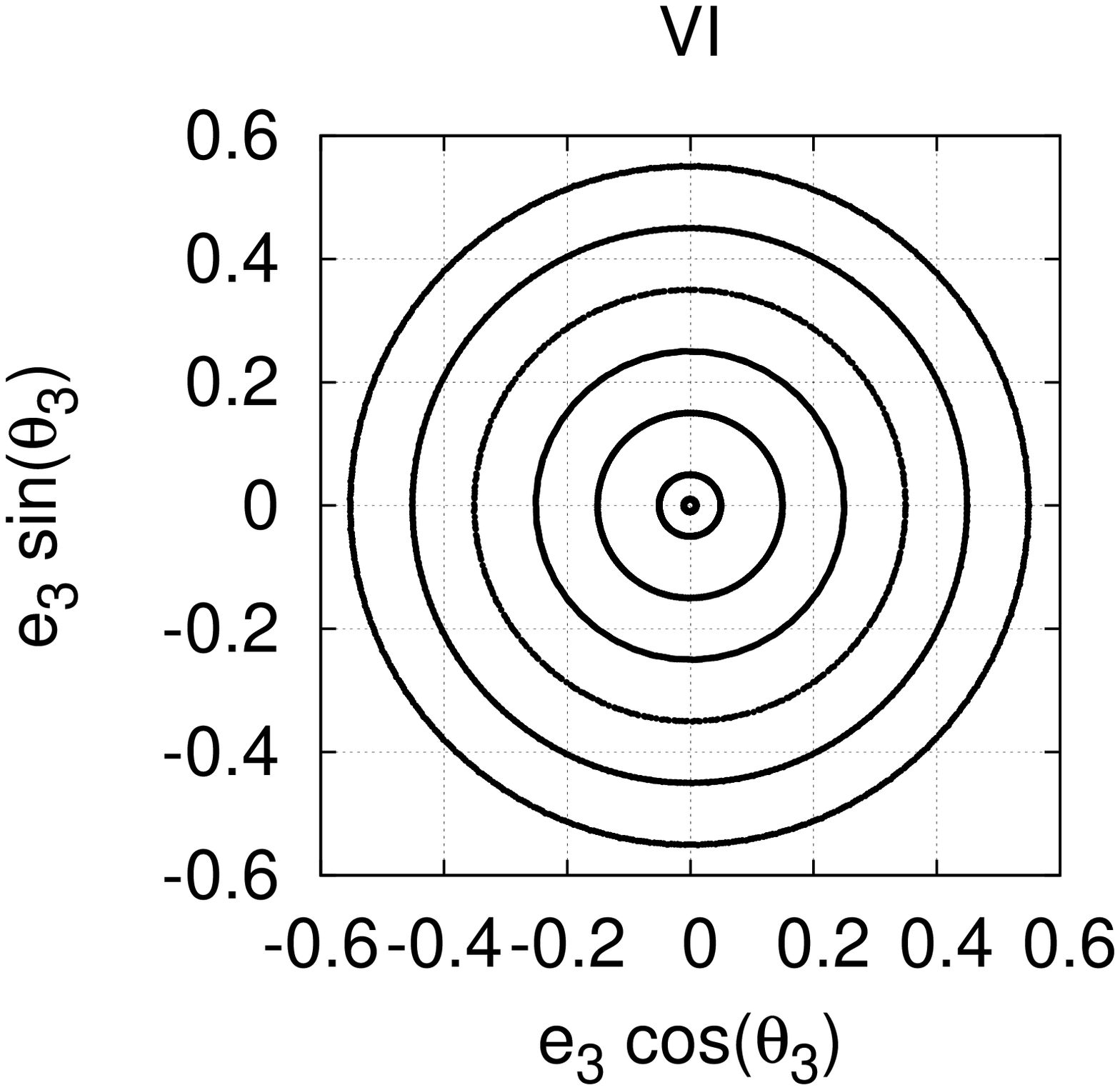}
\caption{Numerical integrations of the N-body problem for different semimajor axis $a_3$ projected on the plane $(e_3 \cos\theta_3,e_3 \sin\theta_3 )$. See the text for more details.} 
\label{levels}
\end{center}
\end{figure}

The forced eccentricity of the second planet being so different along the different branches, it is natural to expect that this effect is noticeable in numerical integrations of the exact equations of motion. Therefore, to measure the influence of the secular resonance in the complete problem, we integrated the complete N-body problem for the same system. For six different values of $a_3$, labelled from I to VI, we plot the projection on the plane $(e_3 \cos\theta_3,e_3 \sin\theta_3)$ of several trajectories resulting from initial conditions selected in this plane and with the initial values for the first planet variables set at $e_2=0.032$ and $\theta_2=0$. The corresponding phase portraits are displayed in Figure \ref{levels}.

Figure \ref{levels} shows that the predictions for the forced eccentricity of the fixed points obtained by the model (Figure \ref{fixed_point-gs-HO-p3}) are remarkably accurate. The panel I of Figure \ref{levels} corresponds to a single fixed point located close to the origin, for a system which semimajor axis ratio $a_2/a_3$ is still large enough to have any influence of the secular resonance. As we decrease the semimajor axis ratio $a_2/a_3$, we approach the secular resonance and even though there is no bifurcation yet, the panels II and III clearly show a displacement of the fixed point reflected by the increase of its forced eccentricity.

Further decreasing the semimajor axis ratio $a_2/a_3$ we get to the panel IV, where we can clearly see the effects of the bifurcation, with two stable fixed points, at $(e_3,\theta_3)\approx (0.25,0)$ and $(e_3,\theta_3)\approx (0.05,\pi)$, and a unstable one, at $(e_3,\theta_3)\approx (0.22,\pi)$. These fixed points correspond to the branches (b), (c) and (d) of Figure \ref{fixed_point-gs-HO-p3}, respectively. In panel V we note that the region where the secular orbits oscillate around the branch (c) increases with the decreasing the semimajor axis ratio $a_2/a_3$, as the forced eccentricity of the branches (b) and (d) increase. Finally, the forced eccentricities of the branches (b) and (d) are high enough such that their orbits become unstable in the complete integrations, leaving us with the motion only around the branch (c) at the panel VI.

\section{Post-Newtonian Effects}
\label{sec:PPN}

The large masses of the stellar components and how close they can be in the circumbinary configuration may have significant relativistic effects that may change the dynamics of the secular motion. Particularly, the post-Newtonian correction introduces a precession in the binary star, that change the value of the secular frequency. In the reference frame adopted, the Post-Newtonian secular Hamiltonian of a 2-body coplanar system is given by (Naoz et al., 2013)
\begin{equation}
{\widetilde H}_{PN} = -\frac{3 \mathcal{G}^2 m_0 m_1 m}{c^2 a_1^2 \sqrt{1-e_1^2}} + {\cal O}(c^{-4}),
\label{PN-eq:01}
\end{equation}
where $c$ is the speed of light. Using (\ref{eq:red_eb}) to express $e_1$ en terms of $e_b$ and $y_j$ at (\ref{PN-eq:01}), we get 

\begin{equation}
\begin{split}
{\widetilde H}_{PN}  & = \frac{3 n_1 \mathcal{G} m}{c^2 a_1 (1-e_b^2)} \left( \sum_{j=2}^{N} |y_j|^2 \right)  + \gO\left(c^{-4}, \vert y_j\vert^4\right)  \\
                                 & = g_{PN} \left( \sum_{j=2}^{N} |y_j|^2 \right)  + \gO\left(c^{-4}, \vert y_j\vert^4\right) .
\end{split}
\label{PN-freq}
\end{equation}

Note that, up to the quadratic terms, the post-Newtonian Hamiltonian introduces only a correction to the secular frequency (and consequently the forced eccentricity) of the planets in the system. In the frame of the quadratic approximation (Section \ref{sec:quadratic}), the corrected frequency is obtained simply by replacing the frequencies 
\begin{equation}
g_j \to g_j - g_{PN}.
\end{equation}
Naturally, this may shift the location of the secular resonance as it was presented in Section \ref{sec:sec_res}. Moreover, we note from (\ref{PN-freq}) that the post-Newtonian is independent of the parameters of the planets, and therefore, this resonance may occur even in the single planet scenario. This effect has already been proposed by many authors (\emph{e.g.,} Ford et al. 2000, Naoz et al. 2013, among others).

\subsection{A single planet}

In the case of a single planet, the secular post-Newtonian frequencies are given by (see Eq. \ref{eq:freq_single})
\begin{equation}
g_2 = - (B_2^{(2)} + \eps R^{(2)}) - g_{PN} = 0.
\end{equation}
Within the domain of validity of the development, $R^{(2)}$ will never be of the order of $B_2^{(2)}$ and, therefore, may be neglected as a first approximation. The resonant condition for the post-Newtonian secular problem is characterized when the frequency $g_2$ vanishes, and that occurs when $\alpha_2$ reaches a critical value $\alpha_{2PN}$ given by
\be
\alpha_{2PN}^{7/2} = \frac{4 G m^3}{a_1 c^2 m_0 m_1} \left(1 + \frac{1}{2}e_b^2 - \frac{3}{2}e_b^4 \right)^{-1}.
\label{eq:cond_RPN3B}
\ee
\subsection{Two planets (n=3)}

In the two planets scenario, the frequencies of the planets are given by (see Eq. \ref{eq:eigenvalues})
\be
g_j =  -\left( B_j^{(2)} + g_{PN} + \eps P^{(2)}_j + \eps R^{(2)} \right)+ \gO (\eps^2){\rm ,\ } j \in \{2,3\}.
\ee
We note that, as $g_{PN}$ is independent of the planetary parameters, it plays a role of a free parameter in the calculation of the secular frequencies. 

The secular resonance  we focused on in the Newtonian case was defined by $g_3= 0$ where the associated $\alpha_3$ satisfied the equation (\ref{eq:g3-resso}).  
If the relativistic precession frequency of the binary $g_{PN}$ is small with respect to the frequency $g^{(r)} = \eps R^{(2)}$  generated by the planets,  the location of the secular resonance, now given by 
$$
B_3^{(2)} + g_{PN} + \eps P^{(2)}_3 + \eps R^{(2)} =0 ,
$$
 is only slightly modified.  This is no longer true when the relativistic precession is dominant, where even the resonance $g_2=0$ can be reached.   
 Therefore, in the general case, there are two possible resonant conditions that may occur in the two planets circumbinary post-Newtonian secular problem, namely
%
%
\be
B_j^{(2)} + g_{PN} + \eps P^{(2)}_j + \eps R^{(2)} = 0 {\rm ,\ }\quad  j\in \{2,3\}.
\label{eq:cond_RPN4B}
\ee
The resonant condition is a polynomial equation on $\alpha_j$ that does not have an analytical solution. As $g_{PN}$ acts as a free parameter, there are combinations of $B_j^{(2)}$,  $P^{(2)}_j$ and $R^{(2)}$ that either of them may play an important role in the location of the resonance. In our particular problem, we chose to solve this equation numerically for each case.

\subsection{Influence of the post-Newtonian correction}

In this section we estimate the influence of the post-Newtonian correction in the known circumbinary planetary systems presented in Table \ref{parameters}, considering a second planet of mass $m_3=10^{-4} M_\odot$ on each system. We present at Table \ref{dynamical_parameters} the semimajor axis of the putative planet at the secular resonance obtained from the three different approaches discussed in this paper, following the notation:

\begin{itemize}
\item $R$ stands for the model that considers only Newtonian interactions (obtained from Eq. \ref{eq:alp-SR});
\item $PN$ stands for the model that considers the post-Newtonian interactions and neglects the influence of the secondary planet (obtained from Eq. \ref{eq:cond_RPN3B});
\item $RPN$ stands for the model that considers both the post-Newtonian interactions and the influence of the secondary planet (obtained from Eq. \ref{eq:cond_RPN4B});
\end{itemize}
 while the two indicators $\sigma_R = \vert a_{3RPN} - a_{3R}\vert / a_{3RPN}$ and $\sigma_{PN} = \vert a_{3RPN} - a_{3PN}\vert / a_{3RPN}$, give an idea of the accuracy of the corresponding model.

\begin{table}
\caption{Dynamical parameters obtained by the models for the systems given in Table \ref{parameters} with the Newtonian frequency of the binary due to the first planet ($\eps R^{(2)}$) and due to the post-Newtonian correction ($g_{PN}$). We present as well the position of the Secular resonance for the 4-body problem $a_{3R}$, the position of the Post-Newtonian Secular Resonance $a_{3PN}$ and the combined effects of the Post-Newtonian and the classical 4-body problem $a_{3RPN}$. In the last two columns we present the error of classical problem ($\sigma_R$) and of the Post-Newtonian ($\sigma_{PN}$), both relative to the result of the combined effect.}
\label{dynamical_parameters}
\begin{tabular}{lllllllll}
\hline
System  &  $\eps R^{(2)}$  &  $g_{PN}$  &  $a_{3R}$  &  $a_{3PN}$  &  $a_{3RPN}$  &  $\sigma_R$  &  $\sigma_{PN}$ \\
  &  $rad/yr$  &  $rad/yr$  &   $[au]$  &   $[au]$  &  $[au]$   &  \%  &  \% \\ \hline
Kep-16b  & $ 4.41\times 10^{-4} $ & $ 6.65\times 10^{-5} $ &  3.64  &  12.0  &  3.63  &  0.41  &  231.7 \\
Kep-34b  & $ 4.99\times 10^{-5} $ & $ 2.62\times 10^{-5} $ &  9.34  &  10.7  &  8.14  &  14.80  &  31.9 \\
Kep-35b  & $ 1.48\times 10^{-4} $ & $ 3.19\times 10^{-5} $ &  5.24  &  8.11  &  4.95  &  5.75  &  63.6 \\
Kep-38b  & $ 8.77\times 10^{-4} $ & $ 2.96\times 10^{-5} $ &  2.40  &  6.30  &  2.36  &  1.52  &  166.3 \\
Kep-47b  & $ 1.10\times 10^{-4} $ & $ 1.52\times 10^{-4} $ &  3.34  &  3.05  &  2.61  &  27.97  &  17.0 \\
Kep-47c  & $ 6.85\times 10^{-5} $ & $ 1.52\times 10^{-4} $ &  7.4  &  3.05  &  3.0  &  145.44  &  1.3 \\
Kep-64b  & $ 4.82\times 10^{-4} $ & $ 3.97\times 10^{-5} $ &  3.36  &  6.81  &  3.28  &  2.31  &  107.3 \\
Kep-413b & $ 5.62\times 10^{-4} $ & $ 9.60\times 10^{-5} $ &  2.45  &  4.06  &  2.34  &  4.48  &  73.1 \\
Kep-1647b & $ 1.06\times 10^{-5} $ & $ 1.00\times 10^{-4} $ &  9.7  &  5.09  &  4.95  &  95.96  &  2.8 \\
\end{tabular}
\end{table}

Note that for the majority of the systems depicted in Table (\ref{dynamical_parameters}) the post-Newtonian interaction is negligible in comparison to the Newtonian interactions due to the planet. This is the case for the systems Kep-16b, Kep-35b, Kep-38b, Kep-64b and Kep-413b, where we have $\eps R^{(2)} \gg g_{PN}$. For these systems, the location and dynamics of the secular resonance are very closely described only by the Newtonian model, as it is also indicated by $\sigma_R < 6\%$.

There are some systems, however, that either the post-Newtonian interactions are very strong or the Newtonian perturbations of the inner planet are very weak, in a way that we have the opposite scenario. For the systems  Kep-47c and Kep-1647b we have $\eps R^{(2)} \ll g_{PN}$, that indicates that the secular resonance can be approximately described solely by the post-Newtonian model. 

Interestingly enough, there are two systems, namely Kep-34b and Kep-47b, that both frequencies $\eps R^{(2)}$ and $g_{PN}$ are of the same order. For both of these systems, we note that neglecting either contribution will lead to significant errors in the location of the secular resonance, as $\sigma_{R},\sigma_{PN} > 15\%$, which means that a complete and more complex model must be adopted. 

\section{Conclusions}
\label{sec:conclusions}

In this work we presented a general formalism for an analytical approach to study the secular dynamics of systems with $N-2$ planets orbiting a circumbinary system. The analytical development presented combines elements from both Legendre polynomials and Laplace coefficients, which allowed us to consider systems with highly eccentric binaries and planetary orbits close to one another. We showed that this development wields similar results when compared with classical models for the case with a single planet, while also providing information regarding the precession of the binary in the case the planetary mass is not negligible.

We calculated the forced eccentricity of the fixed points and the secular frequencies of a system consisting of two planets orbiting a binary star system and we showed that a secular resonance may occur. In this configuration, the inner planet accelerates the precession of the binary, which allows it to enter in resonance with the outer planet. With a simplified model, we presented an expression that gives the approximate location of said resonance and we showed that, for some known systems, that this resonance can occur as close as 2.4 au (for the Kepler-38 system, see Table \ref{parameters}).

With a more refined model, we presented the results obtained for a more thorough development to higher semimajor axis ratios and eccentricities to further study the dynamics of the secular resonance. With this more accurate model, we showed that the secular resonance is associated to a bifurcation of the fixed points into two branches, one being elliptical (stable) and the other, hyperbolic (unstable). This may have important implications on a migrating planet crossing the separatrix of the secular resonance which may greatly affect its eccentricity. Therefore, it would be of interest to introduce a migration mechanism, such as a dissipative disk, to study how this resonance can affect the future planetary formation and evolution in such systems.

Finally, we briefly discussed the influence of the post-Newtonian interactions between the binary star in the secular dynamics of the system. We showed that, up to the quadratic approximation, the post-Newtonian term accelerates the precession of the binary and this additional term allows the resonance to occur in closer orbits and even in systems composed by a single planet. We showed that for the majority of the known systems, however, the influence of the post-Newtonian interactions may be negligible, even though that for some of them it can dominate the dynamical landscape of the secular evolution.

\end{document}